\documentclass[twocolumn, twocolappendix]{aastex631}
\usepackage{float}
\usepackage{wasysym}
\usepackage{booktabs}
\usepackage{changepage}
\usepackage{xspace}
\usepackage[encapsulated]{CJK}

\newcommand{\ruby}{RUBIES-UDS-QG-z7\xspace}

\newcommand{\Oiii}{[O\,{\sc iii}]\xspace}

\newcommand{\Msun}{\rm M_\odot}

\newcommand{\re}{R_{\rm e}}
\newcommand{\prospector}{\texttt{Prospector}\xspace}

\begin{document}

\begin{abstract}

We report the spectroscopic discovery of a massive quiescent galaxy at $z_{\rm spec}=7.29\pm0.01$, just $\sim700\,$Myr after the Big Bang. \ruby was selected from public JWST/NIRCam and MIRI imaging from the PRIMER survey and observed with JWST/NIRSpec as part of RUBIES. The NIRSpec/PRISM spectrum reveals one of the strongest Balmer breaks observed thus far at $z>6$, no emission lines, but tentative Balmer and Ca absorption features, as well as a Lyman break. 
Simultaneous modeling of the NIRSpec/PRISM spectrum and NIRCam and MIRI photometry (spanning $0.9-18\,\micron$) shows that the galaxy formed a stellar mass of log$(M_*/M_\odot)=10.23^{+0.04}_{-0.04}$ before $z\sim8$, and ceased forming stars $50-100\,$Myr prior to the time of observation, resulting in $\log(\rm{sSFR/Gyr}^{-1})<-1$. We measure a small physical size of $209_{-24}^{+33}\,{\rm pc}$, which implies a high stellar mass surface density within the effective radius of $\log(\Sigma_{*,\rm e}/\Msun\,kpc^{-2})=10.85_{-0.12}^{+0.11}$ comparable to the highest densities measured in quiescent galaxies at $z\sim2-5$. The 3D stellar mass density profile of \ruby is remarkably similar to the central densities of local massive ellipticals, suggesting that at least some of their cores may have already been in place at $z>7$. 
The discovery of \ruby has strong implications for galaxy formation models: the estimated number density of quiescent galaxies at $z\sim7$ is $>100\times$ larger than predicted from any model to date, indicating that quiescent galaxies have formed earlier than previously expected.

\end{abstract}


\title{RUBIES Reveals a Massive Quiescent Galaxy at z=7.3}

\author[0000-0001-8928-4465]{Andrea Weibel}
\affiliation{Department of Astronomy, University of Geneva, Chemin Pegasi 51, 1290 Versoix, Switzerland}

\author[0000-0002-2380-9801]{Anna de Graaff}
\affiliation{Max-Planck-Institut f\"ur Astronomie, K\"onigstuhl 17, D-69117, Heidelberg, Germany}

\author[0000-0003-4075-7393]{David J. Setton}\thanks{Brinson Prize Fellow}
\affiliation{Department of Astrophysical Sciences, Princeton University, 4 Ivy Lane, Princeton, NJ 08544, USA}

\author[0000-0001-8367-6265]{Tim B. Miller}
\affiliation{Center for Interdisciplinary Exploration and Research in Astrophysics (CIERA), Northwestern University,1800 Sherman Ave, Evanston, IL 60201, USA}

\author[0000-0001-5851-6649]{Pascal A.\ Oesch}
\affiliation{Department of Astronomy, University of Geneva, Chemin Pegasi 51, 1290 Versoix, Switzerland}
\affiliation{Cosmic Dawn Center (DAWN), Denmark}
\affiliation{Niels Bohr Institute, University of Copenhagen, Jagtvej 128, K{\o}benhavn N, DK-2200, Denmark}

\author[0000-0003-2680-005X]{Gabriel Brammer}
\affiliation{Cosmic Dawn Center (DAWN), Denmark}
\affiliation{Niels Bohr Institute, University of Copenhagen, Jagtvej 128,
K{\o}benhavn N, DK-2200, Denmark}

\author[0000-0003-3021-8564]{Claudia D.P. Lagos}
\affiliation{International Centre for Radio Astronomy Research (ICRAR), M468, University of Western Australia, 35 Stirling Hwy, Crawley, WA 6009, Australia}
\affiliation{ARC Centre of Excellence for All Sky Astrophysics in 3 Dimensions (ASTRO 3D)}
\affiliation{Cosmic Dawn Center (DAWN), Denmark}

\author[0000-0001-7160-3632]{Katherine E. Whitaker}
\affiliation{Department of Astronomy, University of Massachusetts, Amherst, MA 01003, USA}
\affiliation{Cosmic Dawn Center (DAWN), Denmark}

\author[0000-0003-2919-7495]{Christina C.\ Williams}
\affiliation{NSF's National Optical-Infrared Astronomy Research Laboratory, 950 North Cherry Avenue, Tucson, AZ 85719, USA}
\affiliation{Steward Observatory, University of Arizona, 933 North Cherry Avenue, Tucson, AZ 85721, USA}

\author[0009-0005-2295-7246]{Josephine F.W. Baggen}
\affiliation{Department of Astronomy, Yale University, New Haven, CT 06511, USA}

\author[0000-0001-5063-8254]{Rachel Bezanson}
\affiliation{Department of Physics and Astronomy and PITT PACC, University of Pittsburgh, Pittsburgh, PA 15260, USA}

\author[0000-0002-3952-8588]{Leindert A. Boogaard}
\affiliation{Max-Planck-Institut f\"ur Astronomie, K\"onigstuhl 17, D-69117, Heidelberg, Germany}
\affiliation{Leiden Observatory, Leiden University, PO Box 9513, NL-2300 RA Leiden, The Netherlands}

\author[0000-0001-7151-009X]{Nikko J. Cleri}
\affiliation{Department of Astronomy \& Astrophysics, The Pennsylvania State University, University Park, PA 16802, USA}
\affiliation{Institute for Computational \& Data Sciences, The Pennsylvania State University, University Park, PA 16802, USA}
\affiliation{Institute for Gravitation and the Cosmos, The Pennsylvania State University, University Park, PA 16802, USA}

\author[0000-0002-5612-3427]{Jenny E. Greene}
\affiliation{Department of Astrophysical Sciences, Princeton University, 4 Ivy Lane, Princeton, NJ 08544, USA}

\author[0000-0002-3301-3321]{Michaela Hirschmann}
\affiliation{Institute of Physics, Laboratory for galaxy evolution, EPFL, Observatory of Sauverny, Chemin Pegasi 51, 1290 Versoix, Switzerland}

\author[/0000-0002-4684-9005]{Raphael E. Hviding}
\affiliation{Max-Planck-Institut f\"ur Astronomie, K\"onigstuhl 17, D-69117, Heidelberg, Germany}

\author[0009-0009-2845-9255]{Adarsh Kuruvanthodi}
\affiliation{Department of Astronomy, University of Geneva, Chemin Pegasi 51, 1290 Versoix, Switzerland}

\author[0000-0002-2057-5376]{Ivo Labb\'e}
\affiliation{Centre for Astrophysics and Supercomputing, Swinburne University of Technology, Melbourne, VIC 3122, Australia}

\author[0000-0001-6755-1315]{Joel Leja}
\affiliation{Department of Astronomy \& Astrophysics, The Pennsylvania State University, University Park, PA 16802, USA}
\affiliation{Institute for Computational \& Data Sciences, The Pennsylvania State University, University Park, PA 16802, USA}
\affiliation{Institute for Gravitation and the Cosmos, The Pennsylvania State University, University Park, PA 16802, USA}

\author[0000-0003-0695-4414]{Michael V. Maseda}
\affiliation{Department of Astronomy, University of Wisconsin-Madison, 475 N. Charter St., Madison, WI 53706, USA}

\author[0000-0003-2871-127X]{Jorryt Matthee}
\affiliation{Institute of Science and Technology Austria (ISTA), Am Campus 1, 3400 Klosterneuburg, Austria}

\author[0000-0002-2446-8770]{Ian McConachie}
\affiliation{Department of Astronomy, University of Wisconsin-Madison, 475 N. Charter St., Madison, WI 53706, USA}

\author[0000-0003-2895-6218]{Rohan P.\ Naidu}
\altaffiliation{NHFP Hubble Fellow}
\affiliation{MIT Kavli Institute for Astrophysics and Space Research, 77 Massachusetts Ave., Cambridge, MA 02139, USA}

\author[0000-0002-4140-1367]{Guido Roberts-Borsani}
\affiliation{Department of Astronomy, University of Geneva, Chemin Pegasi 51, 1290 Versoix, Switzerland}

\author[0000-0001-7144-7182]{Daniel Schaerer}
\affiliation{Department of Astronomy, University of Geneva, Chemin Pegasi 51, 1290 Versoix, Switzerland}
\affiliation{CNRS, IRAP, 14 Avenue E. Belin, 31400 Toulouse, France}

\author[0000-0002-1714-1905]{Katherine A. Suess}
\affiliation{Department for Astrophysical \& Planetary Science, University of Colorado, Boulder, CO 80309, USA}

\author[0000-0001-6477-4011]{Francesco Valentino}
\affiliation{European Southern Observatory, Karl-Schwarzschild-Str. 2, 85748 Garching, Germany}
\affiliation{Cosmic Dawn Center (DAWN), Denmark}

\author[0000-0002-8282-9888]{Pieter van Dokkum}
\affiliation{Department of Astronomy, Yale University, New Haven, CT 06511, USA}

\author[0000-0001-9269-5046]{Bingjie Wang (\begin{CJK*}{UTF8}{gbsn}王冰洁\ignorespacesafterend\end{CJK*})}
\affiliation{Department of Astronomy \& Astrophysics, The Pennsylvania State University, University Park, PA 16802, USA}
\affiliation{Institute for Computational \& Data Sciences, The Pennsylvania State University, University Park, PA 16802, USA}
\affiliation{Institute for Gravitation and the Cosmos, The Pennsylvania State University, University Park, PA 16802, USA}

\section{Introduction}

The existence of massive quiescent galaxies in the early Universe has posed a longstanding challenge for galaxy formation models, sparking an extensive search for the highest redshift quiescent galaxies \citep[e.g.][]{Marchesini2010,Gobat2012,Glazebrook2017,Schreiber2018b,Santini2019,Carnall20,Forrest2020b,Forrest2020,Valentino2020,Gould23,AntwiDanso2023,Urbano2024}. Although the latest generation of cosmological simulations are able to form massive quiescent galaxies at a rate that matches observations for galaxies with stellar masses $M_*>10^{10}\,\Msun$ out to $z\sim3$, the predicted number density of such systems drops dramatically toward higher stellar masses and redshifts \citep[e.g.][]{Lovell2023,Lagos2024,Kimmig2023,Hartley2023}. However, it is likely that there are quiescent galaxies at even earlier times: approximately half of the compact massive galaxies found at $z\sim2-3$ are already quiescent \citep{Brammer2011,Whitaker2012,vanDokkum2015}, which implies that these systems must have formed a stellar mass $>10^{10}\,\Msun$ \textit{and} ceased forming stars within just $\sim$2-3 Gyr after the Big Bang. The stellar mass densities in the compact quiescent galaxies are also remarkably high, and these systems are therefore thought to evolve into the centers of massive early-type galaxies at the present day \citep{Bezanson2009,vanDokkum2014,Belli2014}.

The cores of massive early-type galaxies in the local Universe thus may have already been in place at $z>3$. 
Indeed, deep spectroscopic studies of massive quiescent galaxies at $z\sim2$ have shown that the stellar populations are very old and strongly $\alpha$-enhanced \citep{Kriek2016,Jafariyazani2020,Beverage2024,Beverage2024b}, indicating that these galaxies formed in a rapid burst of star formation at $z>3$ and possibly as early as $z\sim8$. This is further corroborated by the finding of massive quiescent galaxies at $z\sim3-4$ for which the formation timescales inferred from photometry and/or ground-based spectroscopy imply that some systems formed and quenched already at $z>6$ \citep[e.g.][]{Schreiber2018b,Carnall20,AntwiDanso2023}.
However, prior to quenching, the star-formation histories (SFHs) of these galaxies are notoriously uncertain and sensitive to systematic modeling assumptions. Pinning down their formation histories requires directly observing progenitors at higher redshifts. 

Although compact star-forming galaxies have been found at $z\gtrsim3$ \citep[e.g.][]{Shibuya2015, Bouwens2022}, the search for massive quiescent galaxies at these high redshifts has been limited by the wavelength coverage of the Hubble Space Telescope (HST) and ground-based near-infrared (IR) facilities, which do not extend beyond the Balmer break (rest-frame $\sim 4000\,$\AA) for $z\gtrsim3-4$ \citep[e.g.][]{Glazebrook2017,Schreiber18,Forrest2020,AntwiDanso2023,Tanaka24}. Space-based observations with Spitzer extended the wavelength coverage in the near-IR, aiding the identification of robust quiescent galaxy candidates at $z\sim3-5$ \citep[e.g.][]{Schreiber18,Merlin2018,Merlin19,Carnall20,Gould23}. However, source confusion due to poor spatial resolution and comparatively shallow photometric constraints left the nature of many other candidates ambiguous.

With its high sensitivity, resolution and broad wavelength coverage in the near- and mid-IR, the James Webb Space Telescope (JWST) provides a major leap forward in the search for high-redshift quiescent galaxies. 
Deep near-IR imaging of extragalactic legacy fields with JWST/NIRCam \citep{Rieke2023} has yielded a large number of robust photometric candidates at $z\sim3-5$ \citep[e.g.][]{Carnall2023,Valentino2023,Long2024,Alberts23}. 
Follow-up spectroscopy with JWST/NIRSpec \citep{Jakobsen2022} has confirmed the quiescent nature and high stellar masses for several candidates, with redshifts as high as $z\sim 4.5-5.0$ \citep[][]{Carnall2023b,Carnall2024,Nanayakkara2024,Glazebrook2023,Setton2024,deGraaff2024, Barrufet2024, Wu2024}. 
The discovery of these galaxies continues to challenge galaxy formation models, as the number densities of massive quiescent galaxies in cosmological simulations are up to 2 orders of magnitude below the observational estimates at $z\gtrsim4$ \citep[]{Valentino2023,deGraaff2024,Weller2024,Lagos2024b}.  
Surprisingly, the SFHs inferred from the NIRSpec spectra of some of these early massive quiescent galaxies point to an extremely early burst of star formation ($z>8$) as well as very early quenching ($z>7$). 

It remains unclear whether the corresponding progenitors have already been identified. \citet{Looser2024} and \citet{Strait2023} have reported the discovery of two galaxies at $z\sim5-7$ that show no signs of star formation in the most recent $\sim5-10\,$Myr. Yet, these systems have stellar masses that are over 100 times lower than the massive quiescent galaxies found at $z\sim2-5$, and are likely only temporarily quiescent \citep[e.g.,][]{Dome24}. Other sources at $z\sim7-8$ have been found to have Balmer breaks from both photometry and spectroscopy \citep{Laporte2023,Vikaeus2024,Trussler2024,Witten2024,Kuruvanthodi2024}. However, the measured break strengths are typically small, stellar masses are often low, spectroscopic confirmations remain rare, and the presence of emission lines points to ongoing star formation activity in many of these systems. \citet{Labbe2023}, \citet{Wang2024b} and \citet{Williams2024} demonstrated that there may be very massive galaxies with significant Balmer breaks among the mysterious population of compact red sources dubbed Little Red Dots (LRDs; \citealt{Matthee2024}) at $z\sim7-8$, and full spectral modeling by \citet{Wang2024b} revealed that their SFHs appear similar to those of the massive quiescent galaxies found at $z\sim4-5$. If these sources are indeed confirmed to be massive galaxies, their central stellar densities match those of the compact quiescent systems found at $z\sim2$ \citep{Baggen2023, Baggen2024}. However, the stellar masses of LRDs are still highly uncertain -- at the $\sim 1$ dex level -- due to the likely presence of luminous active galactic nuclei (AGN) that may dominate the spectral energy distribution (SED) at rest-frame optical wavelengths \citep[e.g.,][]{Williams2024,Wang2024b}.

In this paper, we report the discovery of a massive quiescent galaxy at $z=7.3$, \ruby, confirmed spectroscopically with JWST/NIRSpec as part of the Cycle 2 program RUBIES. We describe our photometric selection and spectroscopic observations in Section~\ref{sec:data} and perform stellar population modeling in Section~\ref{sec:prospector} to infer the SFH and stellar population properties. The morphology and estimated stellar mass density profile are presented in Section~\ref{sec:structure}. We place \ruby in the context of the $z\sim7$ galaxy population in Section~\ref{sec:discussion} and discuss its connection to quiescent galaxies at lower redshifts, star-forming progenitors at $z>8$ as well as the lack of massive quiescent galaxies at $z\sim7$ in simulations. A summary and our main conclusions can be found in Section \ref{sec:conclusions}.

Throughout this work we use a flat $\Lambda$CDM cosmology with cosmological parameters from the nine-year Wilkinson Microwave Anisotropy Probe Observations \citep{WMAP9}, $h=0.6932$ and $\Omega_{m,0}$=0.2865. Magnitudes are reported in the AB system. 

\section{Data}
\label{sec:data}

\subsection{Imaging data}
\label{sec:imaging}

The main target of this paper, \ruby\ lies in the UDS field at (R.A., Dec.) = (2:17:43.11, $-$05:06:44.27) and was originally detected in the publicly available JWST/NIRCam data from the Public Release IMaging for Extragalactic Research program (PRIMER; GO-1837; PI J. Dunlop; see, e.g., \citealt{Donnan24}). PRIMER obtained 8 bands of JWST/NIRCam photometry (F090W, F115W, F150W, F200W, F277W, F356W, F410M, and F444W) as well as two MIRI filters (F770W, F1800W). \ruby was identified as a high-priority (`Priority 0') target for spectroscopic follow-up based on its high photometric redshift, bright apparent magnitude ($\rm F444W\approx 24.6$) and the red shape of its spectral energy distribution (SED), which suggested that it was a massive, evolved galaxy at high redshift.

We use the reduced image mosaics of the UDS field from the DAWN JWST Archive (DJA; version 7.2). These data are reduced using the \texttt{grizli} pipeline \citep{grizli} and are drizzled to a pixel scale of $0\farcs04\,{\rm pix}^{-1}$. For more details on the initial data reduction see \citet{Valentino2023}. For the purpose of analyzing the surface brightness profile, we also produce a custom reduction of the F200W filter in the northern half of the UDS-mosaic on a $0\farcs02\,{\rm pix}^{-1}$ pixel scale (see Section~\ref{sec:morphology}).

\subsection{Photometry}
\label{sec:photometry}

Following \citet{Weibel2024}, the NIRCam images are matched to the F444W resolution using empirical point spread function (PSF) models before performing $0\farcs32$ diameter aperture photometry using \texttt{SourceExtractor} \citep{Bertin1996}. Source detection is based on an inverse-variance weighted stack of all the broad-band long-wavelength images (F277W, F356W and F444W) and fluxes are scaled to total based on Kron apertures and an encircled energy correction to account for the flux in the wings of the PSFs.

To obtain MIRI fluxes we measure the flux enclosed in $0\farcs5$ diameter apertures, after subtracting the local background as a sigma-clipped median flux in a $101\times101$ pixels cutout centered on the source. To match to F444W resolution, we multiply the measured flux by the ratio of the encircled energies at $0\farcs5$ diameter of the F444W and the MIRI PSF respectively. This allows us to then scale the fluxes to total in the same way as the NIRCam fluxes.

We note that the MIRI image reduction on the DJA used an older version of the calibration reference files. We therefore finally correct the measured fluxes using the absolute photometric calibration of the latest reference file (version 203; a correction factor of 0.85 and 1.03 for the F770W and F1800W imaging, respectively).

Our photometric measurements in all filters are listed in Table \ref{tab:phot}.

\begin{table}
\centering
\caption{Positions in degrees and NIRCam and MIRI photometric fluxes measured for \ruby\ in nJy.}
\label{tab:phot}
\begin{tabular}{l|l}
\hline
RA & 34.4296173\\[0.1cm]
DEC & -5.1122962\\[0.1cm]
\hline
\multicolumn{2}{c}{NIRCam}\\ \hline
F090W & -3.7$\pm$7.7\\[0.1cm]
F115W & 45.2$\pm$7.9\\[0.1cm]
F150W & 75.7$\pm$6.7\\[0.1cm]
F200W & 108.3$\pm$5.6\\[0.1cm]
F277W & 186.0$\pm$9.3\\[0.1cm]
F356W & 469.6$\pm$23.5\\[0.1cm]
F410M & 522.5$\pm$26.1\\[0.1cm]
F444W & 527.6$\pm$26.4\\\hline
\multicolumn{2}{c}{MIRI}\\ \hline
F770W & 673.7$\pm$94.8\\[0.1cm]
F1800W & 377$\pm$1043\\\hline

\end{tabular}
\end{table}

\subsection{Spectroscopic data}
\label{sec:nirspec}

NIRSpec spectra for \ruby\ were obtained on July 25, 2024 as part of the Red Unknowns: Bright Infrared Extragalactic Survey (RUBIES) (GO-4233; PIs A. de Graaff and G. Brammer, \citealt{deGraaff2024b}). RUBIES is a JWST Cycle 2 program of NIRSpec multi-shutter array (MSA) observations \citep{Ferruit2022} targeting galaxies in the two extragalactic legacy fields CANDELS EGS and UDS  \citep{Candels1,Candels2}. In the UDS field, the RUBIES sources were selected from public JWST/NIRCam imaging obtained by the PRIMER survey.
The MSA exposures consisted of 48 minutes each with the low-resolution PRISM/CLEAR and the medium-resolution G395M/F290LP disperser/filter combinations. A 3-point nodding pattern was used to observe each target in a 1$\times$3 configuration of open microshutters. 

Full details of the NIRSpec data reduction are provided in \citet{Heintz2024} and \citet{deGraaff2024b}. Briefly, we use version 3 of the \texttt{msaexp} \citep{Brammer2022} pipeline. Compared to version 2 of \texttt{msaexp} described in \citet{Heintz2024}, this uses updated reference files for improved flux calibration and custom calibration files for the bar shadow correction, which were constructed from observations of blank sky shutters. 

\begin{figure*}
     \begin{center}
     \includegraphics[width=1.85\columnwidth]{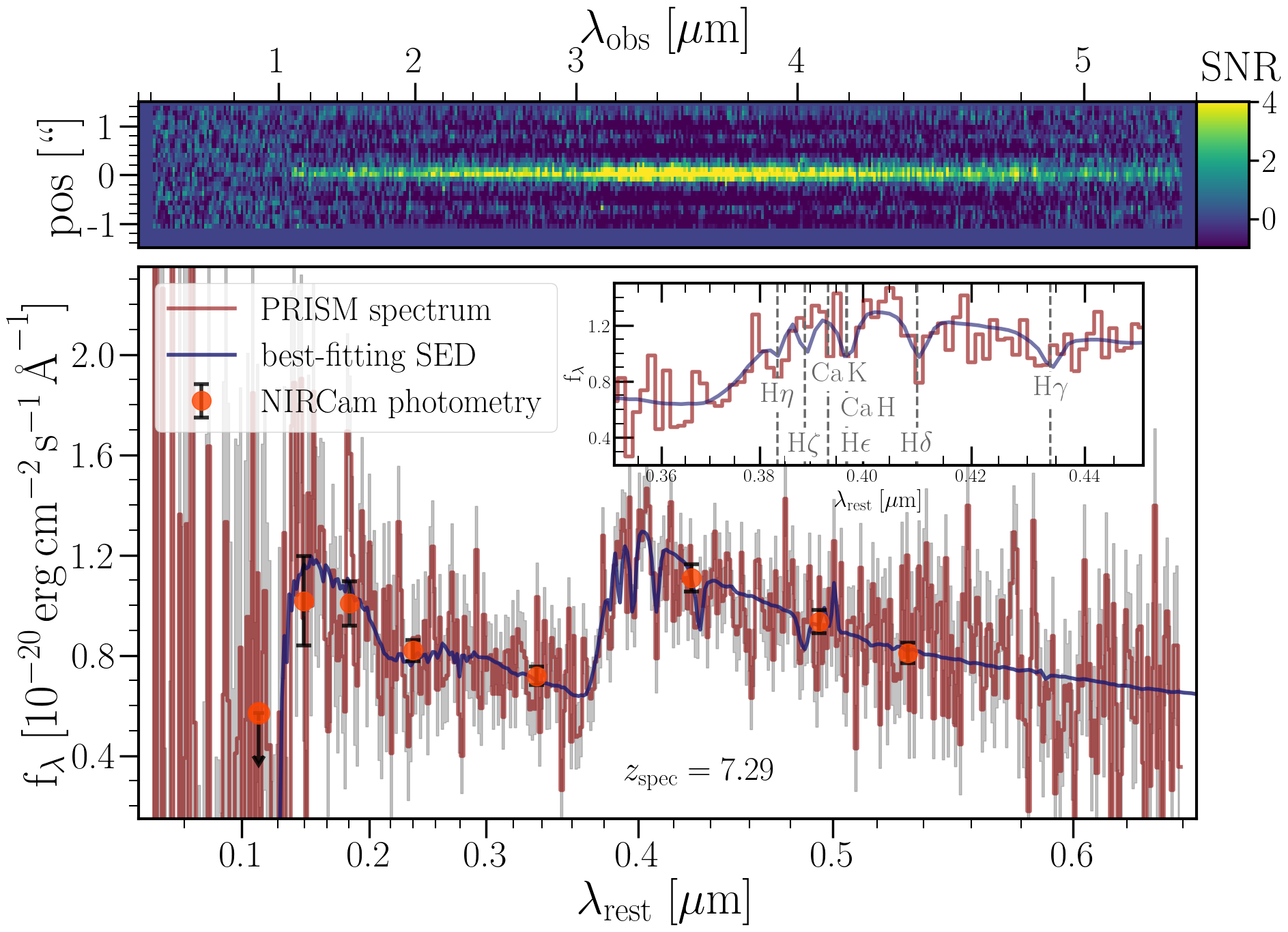}
     \end{center}
     \caption{NIRSpec/PRISM Spectrum of \ruby. Top: 2D SNR spectrum. Bottom: 1D spectrum of \ruby\ in red, with 1$\sigma$ uncertainties in gray. The NIRCam photometry is shown as orange dots and the best-fitting SED from \prospector in blue (see Section \ref{sec:prospector}). The 1D PRISM spectrum has been scaled by the calibration vector inferred by \prospector to account for slit-loss. A zoom-in to the region around $\lambda_{\rm rest}\sim0.4\,\mu$m is shown in the inset panel, where we highlight the position of various absorption features. Note also that the best-fitting SED latches on the Balmer absorption lines.}
     \label{fig:spectrum}
\end{figure*}

We present the PRISM spectrum, both in 2D and in 1D, along with the best-fitting SED (see Section \ref{sec:prospector}) and the NIRCam photometry in Figure \ref{fig:spectrum}. Remarkably, the source shows a clear Balmer break and a Lyman break, unambiguously putting it at a redshift of $z_{\rm spec}=7.29\pm0.01$. There are no emission lines detected at the $>1\sigma$ level. The inset in the bottom panel shows a zoom-in to the region around the Balmer break and highlights the position of various absorption features: The Balmer lines H$\gamma$, H$\delta$, H$\epsilon$, H$\zeta$ and H$\eta$ as well as the Ca H and K lines. While the individual absorption features are observed at a relatively low signal-to-noise ratio (SNR), their combined occurrence makes it unlikely that they are a result of noise.

For completeness, we also show the G395M data in Appendix~\ref{appendix:grating_spectrum}, although the SNR of this spectrum is too low to distinguish any features.

\section{SED modeling}
\label{sec:prospector}

To infer the physical properties of \ruby, we use the Bayesian SED fitting tool \prospector\citep{Johnson2017,Leja2017,Johnson2021} with the nested sampling code \texttt{dynesty} \citep{Speagle2020} to simultaneously fit the PRISM spectrum as well as the NIRCam and MIRI photometry. 

\subsection{\prospector Setup}

We largely follow the same methodology described in detail in  \citet{deGraaff2024}. Briefly, we use the stellar population synthesis models from the Flexible Stellar Population Synthesis (FSPS) package \citep{Conroy2009, Conroy2010}, with the MILES spectral library \citep{Sanchez-Blazquez2006}, MIST isochrones \citep{Choi2016,Dotter2016}, and assuming the initial mass function of \citet{Chabrier2003}. We allow the redshift to vary within $\pm0.1$ around the best-fit redshift of $z=7.288$ measured with \texttt{msaexp}. Due to the comparatively lower signal-to-noise of the spectrum (relative to the massive quiescent galaxy in \citet{deGraaff2024}), we also opt to use a lower order (n=2) polynomial in order to flux calibrate the spectrum to the photometry using the \prospector \texttt{PolySpecFit} procedure. However, our inferred SFHs are insensitive to the particular choice of calibration order.

\begin{figure*}
     \begin{center}
     \includegraphics[width=1.85\columnwidth]{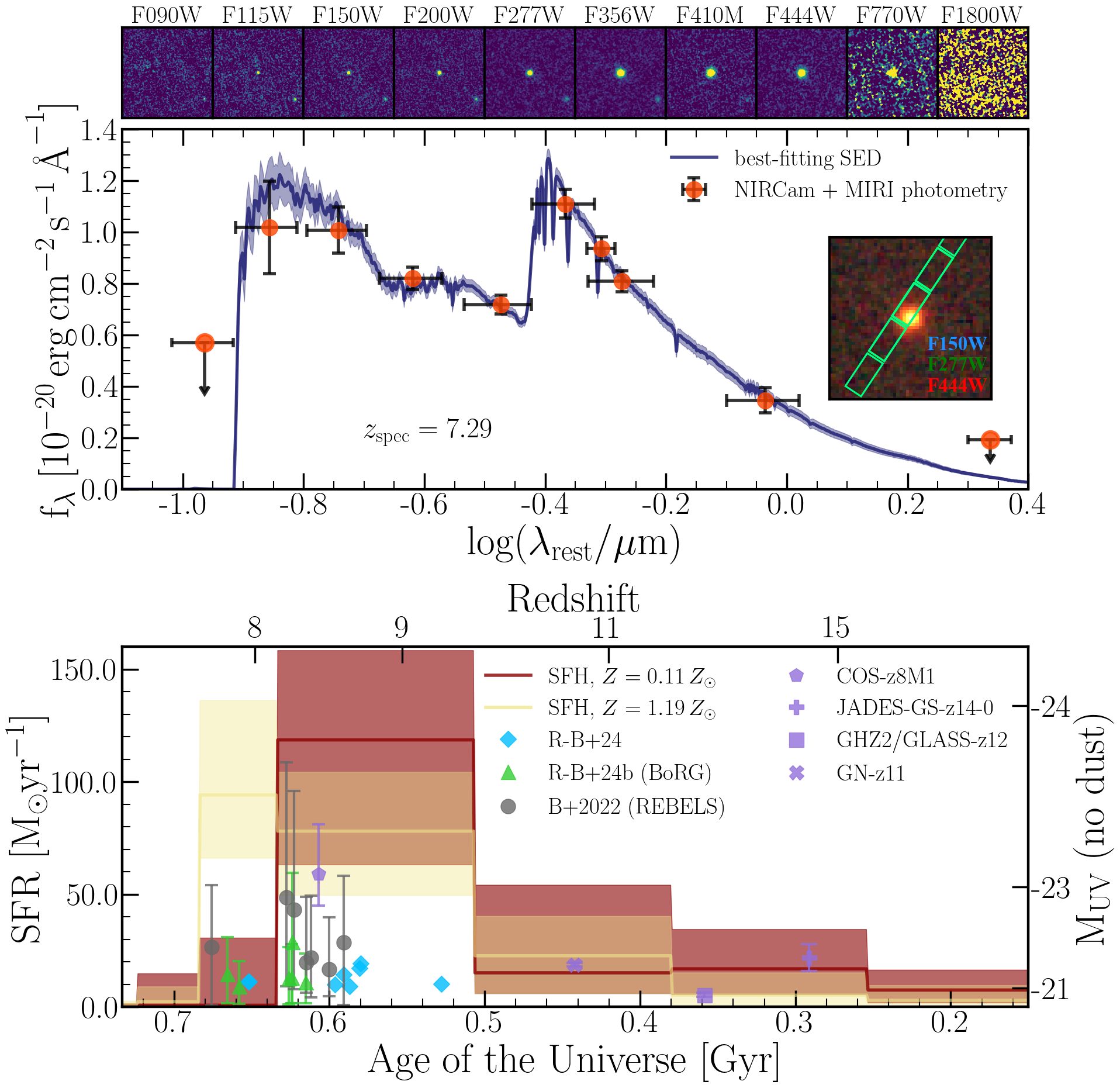}
     \end{center}
     \caption{Top: NIRCam and MIRI imaging cutouts of \ruby. Middle: The posterior median SED from \prospector along with the NIRCam + MIRI photometry, which constrains the fit over the rest-frame optical to near-IR. We further plot the MSA slit position on an inset showing a $2\farcs04\times2\farcs04$ RGB-cutout around the source. Bottom: Fitted non-parametric SFHs for the low-metallicity fit ($Z=0.11\,{\rm Z_\odot}$) in red and the high-metallicity ($Z=1.19\,{\rm Z_\odot}$) fit in khaki. This shows a rising SFH for \ruby\ which peaks at $z\sim8-9$, followed by a rapid decline in the SFR at $z\sim8$ and little to no star formation in the past $\gtrsim50\,$Myr. For comparison, we overplot bright sources at $z\gtrsim8$ with their SFR-values as published in \citet{Akins2023} (COS-z8M1), \citet{Bunker2023b} (GN-z11), \citet{Castellano2024} (GHZ2/GLASS-z12), and \citet{Carniani2024} (JADES-GS-z14-0), as well as three samples of UV-bright objects at $z>7.7$ plotted with their published values of $M_{\rm UV}$, converted to SFR with the relation from \citet{Kennicutt1998}, assuming no dust. Grey: photometrically selected targets for follow-up with ALMA through REBELS from \citet{Bouwens2022b}, light blue: a compilation of NIRSpec observations by \citet{Roberts-Borsani2024}, and light green: objects from the JWST BoRG-survey \citep{Roberts-Borsani2024b}. This illustrates that plausible progenitors of \ruby at $z\sim8-9$, and perhaps at even higher redshifts are either dust-obscured or have yet to be discovered or at least spectroscopically confirmed.}
     \label{fig:sedfit_phot_sfh}
\end{figure*}

We fit a non-parametric SFH that utilizes the continuity prior of \prospector described in \citet{Leja2019}. Due to the higher redshift (and younger age of the Universe at the time of observation) for our source, we adopt a different binning scheme than the one described in \citet{deGraaff2024}. We divide the most recent 100 Myr of cosmic time into 3 bins of 10, 40, and 50 Myr respectively, and fill the remaining earlier cosmic time with 5 linearly spaced bins with widths of $\sim125$ Myr, for a total of 8 bins of star formation.
We assume a two-parameter \citet{Kriek2013} dust law with attenuation around old ($t>10\,$Myr) stars fit in the range $\tau
\in [0, 2.5]$ and a free dust index $\delta \in [-1,0.4]$ that allows for deviations from the \cite{Calzetti2000} dust law and includes a UV bump that depends on the slope parameterized as in \citet{Noll2009}. We fix the attenuation around young ($t<10$ Myr) stars to be twice that of the older populations. We fit the stellar metallicity as a free parameter with a logarithmically sampled uniform prior in the range log($Z/{\rm Z_\odot})\in[-1, 0.19]$. We mask all wavelengths shorter than rest-frame 1200\AA\ to avoid contributions from intergalactic medium absorption. For better visual comparison to the observational data in Figures \ref{fig:spectrum} and \ref{fig:sedfit_phot_sfh}, we apply IGM attenuation to the best-fitting SED using the model from \citet{Inoue2014}. Before fitting, all models are convolved with a line spread function that is a factor 1.3 narrower than the resolution curves available on the JWST User Documentation following e.g., \citet{CurtisLake2023, deGraaff2024} to account for the better resolution of compact sources. We additionally allow for free velocity dispersion parameters that smooth both the stellar continuum and ionized gas emission, that we allow to vary in the range [0,500] km/s to marginalize over the uncertainty in the line spread function in addition to the intrinsic dispersion of the galaxy.

Our initial best-fitting model favors a low metallicity of $Z_{\rm low}=0.11_{-0.01}^{+0.02}\,{\rm Z_\odot}$ which is close to the edge of the prior at $0.1\,{\rm Z_\odot}$. However, we argue that even lower metallicities are not plausible given the high stellar mass of \ruby and considering the observed stellar mass-metallicity relation at these redshifts \citep[e.g.,][]{Nakajima2023,Curti24} or the metallicities of similar-mass and more-massive quiescent galaxies at lower redshift \citep[e.g.,][]{Choi14,Beverage2024}. Since we cannot entirely rule out an even lower metallicity, we discuss this possibility in Section \ref{sec:progenitors}. Further, due to the low resolution and SNR of the PRISM spectrum, the metallicity is only measured indirectly and is sensitive to the detailed shape of the continuum. Instead, the low metallicity may indicate a systematic mismatch between the SPS libraries and the observed SED due to the abundance patterns of stars formed in a rapid burst being $\alpha$-enhanced, as observed in $z=1-3$ quiescent systems \citep{Beverage2024b}. $\alpha$-enhancement can affect the UV continuum shape at the 20-40\% level \citep[e.g.,][]{Vazdekis2015, Choi2019}. Recently, \citet{Park2024} have published novel $\alpha$-enhanced isochrone and spectral libraries, and found similar effects of $\alpha$-enhancement on the continuum. Implementing these new models in SED-fitting codes and exploring their effect on the inferred metallicity and other physical properties will be the subject of future work. Here, we follow an approach similar to the one in \citet{deGraaff2024} (see their Section 4), and attempt to account for this systematic uncertainty due to our inability to marginalize over the effect of varying abundance patterns by running a second fit with a prior on the metallicity of $Z\gtrsim0.3\,{\rm Z_\odot}$, which returns a fit that converges to $Z_{\rm high}=1.19_{-0.27}^{+0.24}\, {\rm Z}_\odot$. 
Comparing the reduced $\chi^2$ of the two fits, we find $\chi^2_{\rm low-Z}=535.05$ and $\chi^2_{\rm high-Z}=538.78$, meaning that we cannot confidently distinguish between the two based on the quality of the fit. The higher metallicity is compensated in the fit by a slightly lower ${\rm A_V}$, marginally lower mass, and a younger age with slightly more star formation in the past $\sim100\,$Myr. The resulting physical parameters for both fits are listed in Table \ref{tab:phys_props}, and the posterior distributions of the most important fitting parameters are shown in Figure \ref{fig:posterior} in Appendix \ref{appendix:posterior}. We emphasize that the low and high metallicity fits are meant to represent the range of plausible metallicities, and likely do not provide meaningful constraints on $Z_*$ individually. However, as can be seen in Figure \ref{fig:posterior}, all the key physical parameters are well constrained independent of $Z_*$.
\\

\subsection{Star Formation History}

In Figure \ref{fig:sedfit_phot_sfh} we show the best-fitting SED from \prospector out to rest-frame near-IR wavelengths, along with the NIRCam+MIRI photometry in the middle panel. The bottom panel shows the inferred non-parametric SFH for the low metallicity and the high metallicity fit.

Both SFHs indicate that \ruby\ forms more than half of its mass in a burst lasting $\sim100-200\,$Myr at around $z\sim8-9$, and with a SFR of $\sim100\,{\rm M}_\odot\,{\rm yr}^{-1}$. Then, the SFR drops to $<3\,{\rm M}_\odot\,{\rm yr}^{-1}$ within a few 10s of Myr around $z\sim8$. The high metallicity fit allows for some more star formation below $z\sim8$, while the low metallicity fit favours slightly earlier quenching and a more extended SFH at $z\gtrsim11$.
Importantly, both SFHs indicate that the galaxy has a stellar mass of $\log(M_*/\rm{M}_\odot)\sim10.2$ and a specific SFR of $\log(\rm{sSFR}/\rm{Gyr^{-1}})\lesssim-1$ when averaged over the past $50\,$Myr.
As a test on the robustness of these properties with regard to changing the SFH-prior, we perform another \prospector run assuming the bursty continuity prior as described in \citet{Tacchella2022}, which allows for more drastic changes in the SFR from one time-bin to the next. This increased flexibility is reflected in larger uncertainties on the SFR in time-bins prior to $50\,$Myr before the time of observation. Crucially, the inferred stellar mass, dust attenuation and SFR$_{50}$ are consistent within errors with those inferred with the continuity prior. In fact, the bursty continuity prior favors an even lower SFR$_{50}$. We further refer the reader to \citet{deGraaff2024} who explored the effect of using parametric forms of the SFH such as a delayed-$\tau$, and a rising SFH model to fit a massive quiescent galaxy at $z\sim4.9$ whose SED resembles that of \ruby (see Figure \ref{fig:spectrum_comparison}), and found no significant changes in the inferred physical parameters.

To get an alternative direct measurement of the upper limit on the recent star formation, we use the python tool \texttt{lmfit} to fit a single Gaussian to the PRISM and G395M spectrum respecitvely within $\pm30\,$\AA\ rest-frame of the H$\beta$ line center at $\lambda_{{\rm H}\beta}=4861\,$\AA\ after subtracting the continuum as the median in that wavelength window. Since the continuum-subtracted spectra are dominated by noise, we enforce a positive amplitude for the fit and constrain the line center to be within $\pm5\,$\AA\ of $\lambda_{{\rm H}\beta}$, as well as the dispersion to be within 200 and 400 km/s to avoid unreasonably wide and/or off-centered fits. The values for the dispersion are based on the relation from \citet{Forrest2022} for the dynamical mass M$_{\rm dyn}$. Assuming M$_{\rm dyn}=M_*$, which can be interpreted as a lower limit on M$_{\rm dyn}$, this yields $\sigma\approx225\,$km/s. From the inferred uncertainty in the fit we get a 2$\sigma$ upper limit on the line flux that is allowed by the data. We then convert the approximate best-fitting dust extinction parameter A$_V=0.3$ to a Balmer decrement to obtain an upper limit on the H$\alpha$ flux which we can finally translate into an upper limit on the recent SFR using the relation from \citet{Kennicutt1998}, obtaining SFR$_{\rm lines}<6.6\,{\rm M_\odot\,yr^{-1}}$ and SFR$_{\rm lines}<5.8\,{\rm M_\odot\,yr^{-1}}$ from the PRISM and G395M spectrum respectively. This is significantly less constraining than the measurement from \prospector of SFR$_{10}=0.64^{+0.83}_{-0.60}\,{\rm M_\odot\,yr^{-1}}$ for the low metallicity model (see Table \ref{tab:phys_props}), due to the fact that the \prospector fit is informed by the absence of multiple lines and the continuum shape of the spectrum. It nevertheless provides another confirmation that the recent star formation rate of \ruby must be low.

The secondary y-axis on the right of the bottom panel of Figure \ref{fig:sedfit_phot_sfh} shows $M_{\rm UV}$ values that are directly inferred from the SFR, assuming the relation from \citet{Kennicutt1998} and no dust extinction. For comparison with the SFH of \ruby, we collect literature sources with the brightest M$_{\rm UV}$ known at $z>7.7$ (i.e., at epochs before \ruby stopped forming stars): a sample of objects selected for follow-up with ALMA through REBELS from \citet{Bouwens2022b}, a compilation of galaxies observed with NIRSpec from \citet{Roberts-Borsani2024} and a sample of sources identified in pure-parallel HST-imaging and spectroscopically confirmed through the BoRG-JWST survey \citep{Roberts-Borsani2024b}. All of these sources remain below the unobscured $M_{\rm UV}\lesssim-23$ suggested by the $\sim50-150\,{\rm M}_\odot\,{\rm yr}^{-1}$ SFR in the SFH of \ruby. We also show a dust-obscured (${\rm A_V}\sim1.6$) galaxy at $z\sim8.4$ from \citet{Akins2023} named COS-z8M1 with an inferred SFR$_{100}\sim59\,{\rm M_\odot}{\rm yr}^{-1}$. Only this object and three sources from \citet{Bouwens2022b} are consistent with the SFH of \ruby within the 1$\sigma$ uncertainties. All of them rely on photometric data only, and the sources from \citet{Bouwens2022b} were selected over an area of as much as $\sim7\,$deg$^2$, whereas the total area covered by RUBIES is only $\sim150$ arcmin$^2$. Finally, we show three remarkably luminous galaxies at $z>10$, GN-z11, GHZ2/GLASS-z12 and JADES-GS-z14-0 with values of SFR$_{10}$ as published in \citet{Bunker2023b}, \citet{Castellano2024}, and \citet{Carniani2024}. While these sources are consistent with potentially being progenitors of \ruby, its SFH is too poorly constrained at $z\gtrsim10$ to draw further conclusions. However, no sources with SFRs as high as the ones predicted during the burst at $z\sim8-9$ have been spectroscopically confirmed to date, with photometric candidates being either dust-obscured and/or discovered in wider area imaging. For a more extended discussion on the possible progenitors of \ruby, see Section \ref{sec:progenitors}.

\begin{table}
\centering
\caption{Physical properties of \ruby, as measured with \prospector for the low metallicity (low-Z) and the high metallicity (high-Z) fit.}
\label{tab:phys_props}
\begin{tabular}{l|l|l}
quantity & low-Z & high-Z\\[0.1cm]\hline
$z_{\rm spec}$ & $7.287^{+0.007}_{-0.006}$& $7.290^{+0.005}_{-0.006}$\\[0.1cm]
log($M_*/{\rm M}_\odot$) & $10.23^{+0.04}_{-0.04}$& $10.19^{+0.04}_{-0.04}$\\[0.1cm]
log($\Sigma_*/{\rm M_\odot}\,{\rm kpc}^{-2}$) & $10.85^{+0.11}_{-0.12}$& $10.80^{+0.11}_{-0.12}$\\[0.1cm]
SFR$_{10}$ $[{\rm M}_\odot\,{\rm yr^{-1}}]$ & $0.64^{+0.83}_{-0.60}$& $1.08^{+1.55}_{-0.98}$\\[0.1cm]
SFR$_{50}$ $[{\rm M}_\odot\,{\rm yr^{-1}}]$ & $0.83^{+11.11}_{-0.76}$& $2.13^{+5.54}_{-1.92}$\\[0.1cm]
SFR$_{100}$ $[{\rm M}_\odot\,{\rm yr^{-1}}]$ & $0.84^{+20.16}_{-0.78}$& $48.89^{+21.12}_{-13.04}$\\[0.1cm]
${\rm A_V}$ [mag] & $0.31^{+0.08}_{-0.08}$& $0.25^{+0.09}_{-0.07}$\\[0.1cm]
$t_{50}$ [Gyr] & $0.20^{+0.07}_{-0.02}$& $0.16^{+0.03}_{-0.02}$\\[0.1cm]
$t_{90}$ [Gyr] & $0.12^{+0.01}_{-0.01}$& $0.07^{+0.01}_{-0.01}$\\[0.1cm]
log($Z/{\rm Z}_\odot)$ & $-0.94^{+0.05}_{-0.04}$& $0.07^{+0.08}_{-0.11}$\\\hline
\end{tabular}
\end{table}

\section{Structural Properties}\label{sec:structure}

To characterize the morphology of \ruby, we first fit a S\'ersic profile to the imaging, and then use this measurement to estimate the stellar mass surface density as well as a 3D density profile.

\subsection{S\'ersic Profile Fitting}
\label{sec:morphology}

We first use the F200W filter to measure the morphology of \ruby, as this is the NIRCam filter with the highest SNR for which imaging at a pixel scale of $0\farcs02$ is available. Size measurements in longer wavelength filters are discussed further below.
We select bright, but not saturated, and isolated stars and use the python tool \texttt{psf.EPSFBuilder} \footnote{\url{https://photutils.readthedocs.io/en/stable/api/photutils.psf.EPSFBuilder.html}} \citep{Anderson2000, Anderson2016} from the \texttt{photutils} package \citep{Bradley2022} to derive an effective PSF. 

Next, we use the empirical PSF to perform single S\'ersic profile fitting with the \texttt{pysersic} package \citep{pysersic}\footnote{\url{https://github.com/pysersic/pysersic}}. To sample the posterior we use the No U-turn sampler implemented in \texttt{numpyro} \citep{hoffman2014, phan2019}. We use two chains with 1000 warm-up and 2000 sampling steps each. This results in effective sample sizes $> 1200$ for all parameters and $\hat{r} < 1.01$ indicating robust sampling \citep{vehtari2021}.

We find that the source is marginally resolved, with a major axis size of approximately 2 pixels, corresponding to a physical size of $\re=209^{+33}_{-24}\,{\rm pc}$. The source is round ($b/a=0.89_{-0.14}^{+0.08}$) and has a poorly constrained S\'ersic index, $n=2.4_{-0.9}^{+1.5}$. Regardless of the uncertain S\'ersic index, the small size implies a high stellar mass surface density within the effective radius of log($\Sigma_{\rm *,e}/{\rm M_\odot}\,{\rm kpc}^{-2})=10.85_{-0.12}^{+0.11}$. 

In Figure \ref{fig:mass_density_profile} (left panel), we compare this density to quiescent galaxies from the literature. First, we compile a sample of 225 quiescent galaxies at $z\sim2-3$ with log$(M_*/{\rm M}_\odot)>10.5$ from the 3DHST catalog \citep{Skelton14}, using the UVJ-color cuts from \citet{Muzzin2013} and matching with the size fits from \citet[][selecting only good fits with \texttt{flag=0}]{vanderWel12} to compute the surface density within the effective radius for each object, similar to the sample in \citet{Whitaker2017}. In addition, we show the surface densities of the massive quiescent galaxy at $z=4.66$ in \citet{Carnall2023b}, of the extremely massive ($M_*=10^{11}\,{\rm M}_\odot$) quiescent galaxy RUBIES-EGS-QG-1 at $z=4.9$ from \citet{deGraaff2024}, of the compact core component of a $z=3.97$ quiescent galaxy from \citet{Setton2024}, and the highest density measured in lensed star clusters at $z\sim6$ by \citet{Vanzella2023}. The surface mass density of \ruby is comparable to those of the densest quiescent galaxies at $z\sim2-3$ as well as the most massive quiescent systems at $z\sim4-5$, and only a factor $\sim4$ less dense than the highest densities measured in star clusters at $z\sim6$.

Following the same methodology, we also fit single S\'ersic profiles to the NIRCam filters in the long wavelength channel, where F356W and F444W probe the rest-frame optical continuum, but both the pixel scale and PSF FWHM are a factor $\sim2$ larger (for reference, we show the PSF half width at half maximum of F200W and F444W in Figure \ref{fig:mass_density_profile}). On the pixel scale of 0\farcs04, we measure sizes of $\lesssim0.5$ pixels, close to the edge of the prior at 0.25 pixels, indicating that \ruby is not resolved in the rest-optical.  Taking the 95th percentiles of the posterior distributions as upper limits on the physical size, we find $\re<103\,$pc and $\re<88\,$pc in F356W and F444W respectively. Under the assumption that our PSF-models are accurate and noise-free, this indicates that \ruby is at least a factor $\times2$ smaller in the rest-optical than in the rest-UV, which is stronger than (but qualitatively consistent with) the color gradients found in massive quiescent galaxies at $z\sim1-4$ \citep[e.g.][]{Suess2019,Suess2022,Cutler2024,Ji2024,Wright2024}. Crucially, this would further imply that the stellar mass density is at least a factor $\times4$ higher than inferred above. 

\subsection{Stellar Mass Density Profiles}
\label{sec:3d_density}

\begin{figure*}
     \begin{center}
     \includegraphics[width=1.85\columnwidth]{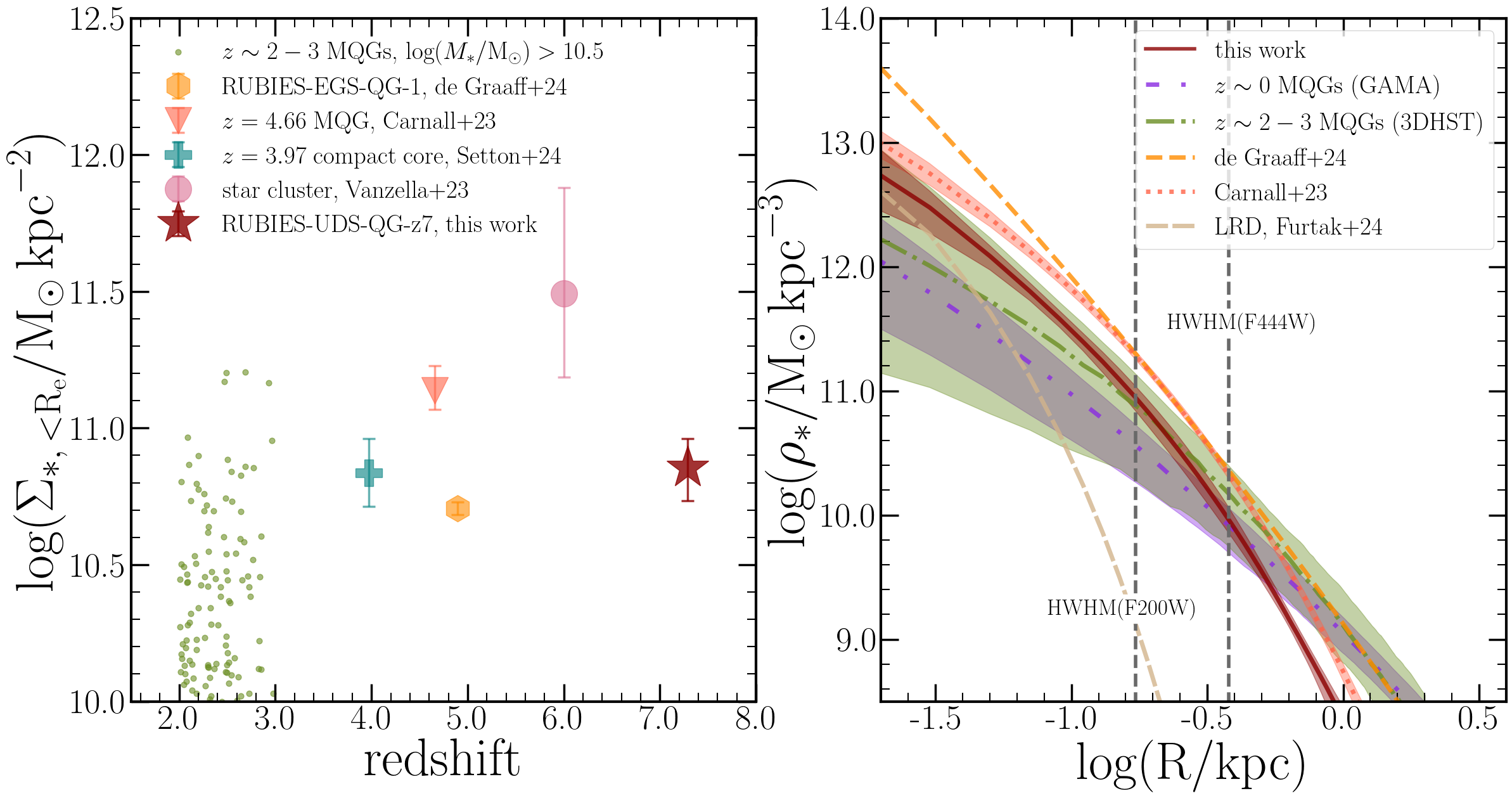}
     \end{center}
     \caption{Left: Projected stellar mass surface density within $\re$ versus redshift for \ruby compared to the massive quiescent galaxy at $z=4.9$ from \citet{deGraaff2024}, a massive quiescent galaxy at $z=4.66$ from \citet{Carnall2023b}, the core component of a massive quiescent galaxy at $z=3.97$ from \citet{Setton2024}, the densest star cluster found in the Sunrise arc by \citet{Vanzella2023}, and a sample of 225 massive quiescent galaxies at $z\sim2-3$ selected from the 3DHST survey. The surface density of \ruby is consistent with the densities of massive quiescent galaxies at $z\sim4-5$ and the densest systems at $z\sim2-3$, and only a factor $\sim4$ below that of the densest known star clusters at $z\sim6$. Right: Mass density profiles of some of the objects shown on the left, as well as of the triply imaged LRD from \citet{Furtak2024}. For the quiescent galaxies at $z\sim2-3$, we show the median stellar mass profile and shade the region between the 16th and 84th percentiles. Similarly, we compile a sample of 514 massive quiescent galaxies at $z\sim0$ from the GAMA survey, and show the respective median and percentile profiles. The stellar mass density of \ruby at $R\sim300\,$pc is consistent with the latter, indicating that the cores of some local ellipticals may be in place at $z\sim7$.
     }
     \label{fig:mass_density_profile}
 \end{figure*}

To estimate the 3D mass profile, we follow the methodology described in \citet{Bezanson2009}.
We perform an Abel transform to deproject the 2D S\'ersic profile measured from the F200W imaging, assuming that the mass density profile is spherically symmetric and that the mass-to-light ratio is constant with radius. As noted above, the assumption of a constant M/L ratio is conservative given the observed F200W and F444W sizes, and may underestimate the true stellar mass density of this object.

The resulting mass density profile of \ruby\ is shown in Figure \ref{fig:mass_density_profile}, and reveals a very high stellar mass density of $\sim 10^{11}\,\Msun\,$kpc$^{-3}$ at the effective radius. For comparison, we also compute the mass density profile of RUBIES-EGS-QG-1 \citep{deGraaff2024}, and of the triply imaged LRD from \citet{Furtak2024}, assuming $\re=30\,{\rm pc}$, $M_*=10^{9.15}\,{\rm M}_\odot$ (corresponding to the upper limit on the stellar mass specified in their paper), and $n=1.5$.
Further, we show the median as well as the 16th and 84th percentile mass density profile of the sample of massive (log$(M_*/{\rm M}_\odot)>10.5$) quiescent galaxies at $z\sim2-3$ described above. In a similar manner, we compile a sample of 514 $z\sim0$ quiescent galaxies from the Galaxy And Mass Assembly survey \citep[GAMA;][]{Driver2011,Liske2015,Baldry2018}, with masses log$(M_*/{\rm M}_\odot)>11$ from \citet{Driver2016}, adopting the selection of quiescent galaxies from \citet{deGraaff2022} and the size fits in the r-band from \citet{Kelvin2012} to compute a median as well as 16th and 84th percentile mass profiles of massive quiescent galaxies in the local Universe.

In comparison to the sources at $z\gtrsim4.5$, we find that the mass density profile of \ruby lies $\sim0.5\,$dex below that of RUBIES-EGS-QG-1, and $0.1-0.2\,$dex below that of the massive quiescent galaxy from \citet{Carnall2023b} at all displayed radii. The triply imaged LRD shows a much more centrally peaked profile, reaching densities comparable to that of \ruby at radii of $\sim10{\rm pc}$, where the mass density profiles are poorly constrained based on the available imaging data.
On the other hand, the mass density profile of \ruby is remarkably consistent with the densest compact quiescent galaxies at $z\sim2-3$, and it lies $\sim0.2\,$dex above the 84th percentile of the local early-type galaxy profiles at radii $\lesssim300\,$pc with some individual $z\sim0$ profiles sill being consistent with the profile of \ruby. Both $z\sim2-3$ and local ellipticals show higher densities at larger radii due to their more extended morphologies which can be explained by the accretion of smaller satellite systems between $0<z<2$ \citep[e.g.][]{Bezanson2009,Naab2009}. As also discussed there, the difference in central densities between massive quiescent galaxies at cosmic noon and local ellipticals is consistent with inside-out growth scenarios, where the $z\sim2-3$ galaxies evolve into the cores of local ellipticals. Critically, this suggests that it is possible for the cores of present-day massive galaxies to already form within the first $700\,$Myr. 

\section{Discussion}
\label{sec:discussion}

\begin{figure*}
     \begin{center}
     \includegraphics[width=1.85\columnwidth]{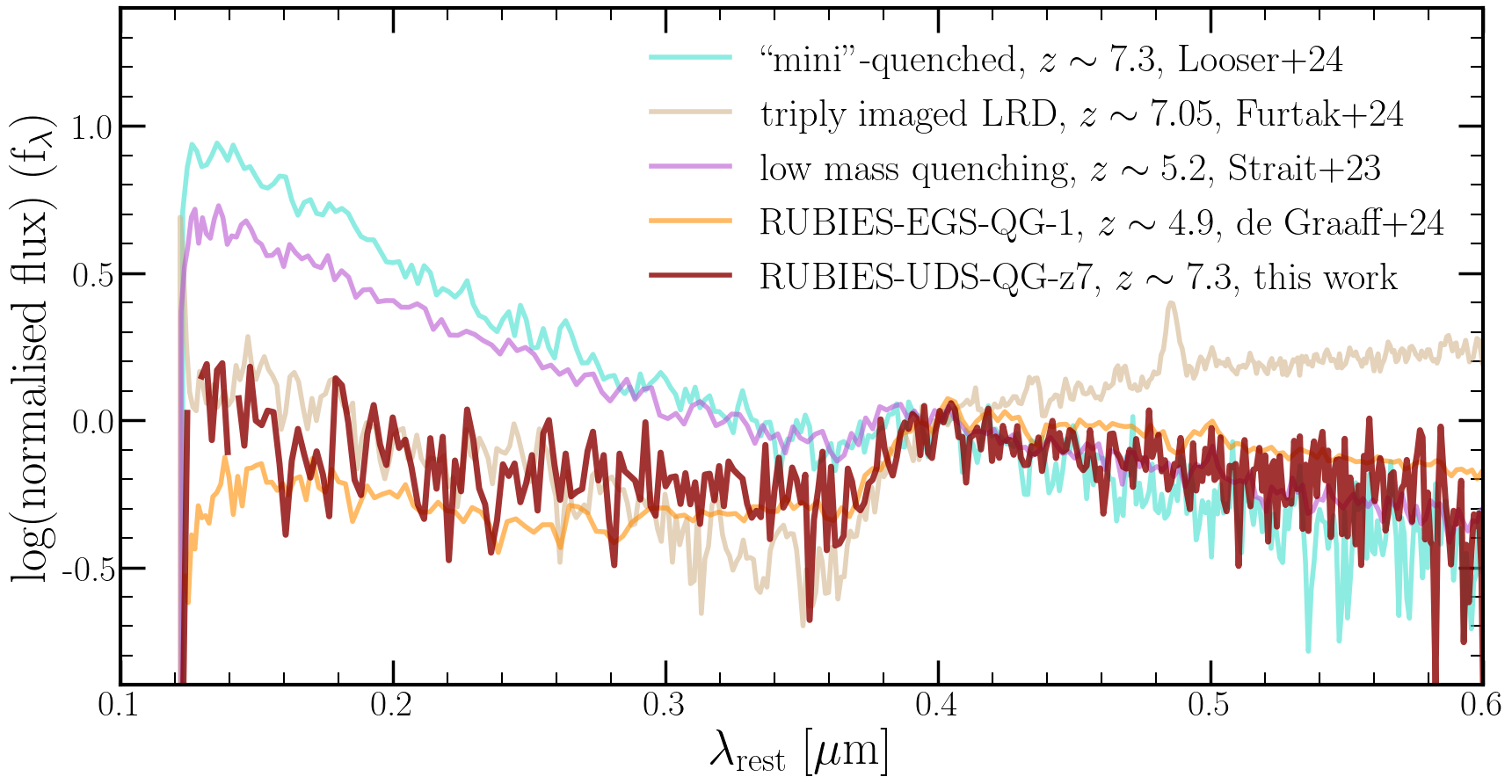}
     \end{center}
     \caption{Comparison of \ruby to various spectra from the literature: the low-mass, recently quenched galaxies (also referred to as (mini-)quenched) at $z\sim7.3$ from \citet{Looser2024} and at $z\sim5.2$ from \citet{Strait2023}, the triply imaged LRD from \citet{Furtak2024}, and the massive quiescent galaxy at $z=4.9$ from \citet{deGraaff2024}. While the former two show much stronger UV-emission than \ruby, the LRD shows a much redder rest-optical continuum and a broad H$\beta$ emission line. For its redshift, the spectrum of \ruby is unique. However, it shows remarkable agreement with the spectral shape of the $z=4.9$ massive quiescent galaxy.}
     \label{fig:spectrum_comparison}
\end{figure*}

Through detailed analysis of the JWST photometric and spectroscopic data, we have established that \ruby has formed a stellar mass $>10^{10}\,{\rm M_\odot}$ by $z\sim8$, retaining a compact morphology with $\re\sim200\,{\rm pc}$, and then stopped forming stars rapidly within a few 10s of Myr and remained quiescent with log(${\rm sSFR/Gyr}^{-1})\lesssim-1$ for the past $\sim50-100\,$Myr. This makes \ruby the highest redshift bonafide massive quiescent galaxy confirmed to date, and raises major questions: How was this galaxy able to form and quench in just $700\,$Myr? What would its likely progenitors and descendants look like? Below, we compare the spectral shape of \ruby to other red objects at similar redshift, place \ruby in the context of galaxy formation models as well as the general galaxy population at $z\sim7$, and discuss its possible past and future.

\subsection{A Unique, Quiescent Galaxy at $z\sim7$}
\label{sec:context}

To put \ruby in context, we compare it against new classes of red and/or (mini-)quenched objects discovered recently with JWST in Figure \ref{fig:spectrum_comparison}. For illustrative purposes, we have normalized the spectra to the median flux at $\lambda_{\rm rest}\in(0.39,0.41)\mu m$. First, we show the ``(mini-)quenched'' object from \citet{Looser2024} (which coincidentally lies at the exact same redshift as \ruby, $z=7.29$, but in the GOODS-S field), as well as a similar object presented in \citet{Strait2023} at $z=5.2$. Both sources show a significantly weaker Balmer Break and, crucially, a very steep (blue) rest-UV slope, indicative of more recent star-formation than in \ruby, which shows $\beta = -0.84\pm0.15$.
Moreover, we find that the flux density at $\sim0.4\,\mu m$ rest-frame of the \citet{Looser2023} source is lower by a factor $\sim12$. Combined, this clearly distinguishes \ruby from these objects that have lower stellar masses ($\sim10^{7.6-8.7}\,\Msun$), and stopped forming stars more recently ($\lesssim 10-30\,$Myr).

Next, we compare with the triply imaged LRD from \citet{Furtak2024} at $z=7.05$. With a ${\rm F115W}-{\rm F200W}$ color of 0.95 mag, reflecting a relatively red rest-UV spectrum, and ${\rm F277W}-{\rm F444W}=1.13$, \ruby lies just inside the color selection boxes proposed for LRDs by \citet{Labbe2023b} and \citet{Greene2024}. We specifically choose the LRD of \citet{Furtak2024} for comparison, because it lies at a similar redshift, has a very high quality spectrum thanks to its lensing magnification, and it does not have strong \Oiii emission \citep[which many other LRDs do; see, e.g.,][]{Greene2024}.  While its spectrum matches that of \ruby well up to $\lambda_{\rm rest}=0.4\mu m$, including a strong Balmer break, the spectrum of the LRD continues to rise towards longer wavelengths and shows a broad H$\beta$ emission line. Both of these features may be attributed to a dust-obscured AGN that dominates the SED at rest-frame optical wavelengths. However, as can be clearly seen in Figure \ref{fig:spectrum_comparison}, \ruby lacks the characteristic rising red continuum of LRDs and its red color between F277W and F356W is solely due to the Balmer Break, while ${\rm F356W}-{\rm F444W}=0.13$ indicates a relatively flat optical continuum in $f_\nu$.

Finally, the spectrum of the massive quiescent galaxy RUBIES-EGS-QG-1 at $z=4.9$ \citep{deGraaff2024} matches that of \ruby remarkably well, corroborating the interpretation that \ruby is truly a massive quiescent galaxy at $z=7.3$.

\begin{figure}[t!]
     \centering
     \includegraphics[width=0.47\textwidth]{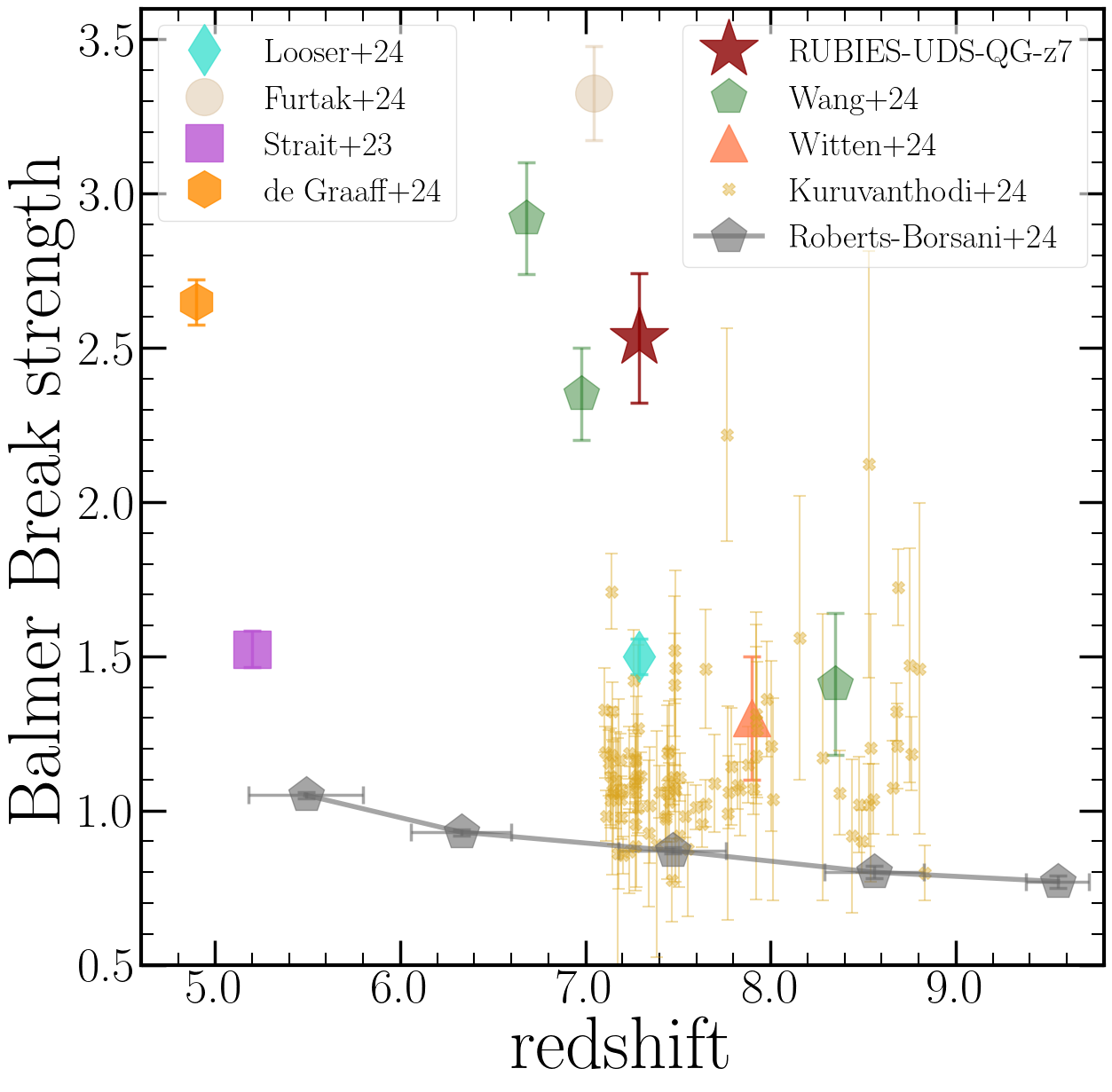}
     \caption{Balmer Break strength vs. redshift for the 5 sources shown in Figure \ref{fig:spectrum_comparison}, three objects with strong breaks from \citet{Wang2024b}, YD4 as published in \citet{Witten2024}, the median relation from stacked NIRSpec spectra from \citet{Roberts-Borsani2024}, and a sample of objects with NIRSpec spectroscopic redshifts and breaks measured from photometry \citep{Kuruvanthodi2024}. \ruby\ shows one of the strongest Balmer breaks measured at $z\gtrsim7$ to date, comparable to the break strength measured in the massive quiescent galaxy at $z=4.9$ \citep{deGraaff2024}.
     }
     \label{fig:bbs}
 \end{figure}

To further highlight the uniqueness of \ruby, we quantify the strength of its Balmer break, which roughly traces the age of the underlying stellar populations or the time since quenching \citep[e.g.,][]{Hamilton85,Kauffmann03,Kriek06}. Using the wavelength windows defined in \citet{Wang2024b} that avoid prominent nebular emission lines around rest-frame 4000\,\AA\, we measure the break strengths of all five objects shown in Figure \ref{fig:spectrum_comparison}, as well as the three objects presented in \citet{Wang2024b}. These measurements are performed consistently on v3 spectra from the DJA, and in f$_\nu$, to be consistent with historical definitions of the break strength \citep[e.g.][]{Bruzual1983,Balogh1999}. A compilation of break strengths is shown in Figure \ref{fig:bbs}, including the object YD4 in the A2744 cluster whose break strength has recently been published by \citet{Witten2024}. For reference, we further plot the median relation from stacked NIRSpec spectra from \citet{Roberts-Borsani2024} and a sample of sources with NIRSpec spectroscopic redshifts compiled by \citet{Kuruvanthodi2024} who measured the Balmer breaks from photometry. While their choice of photometric filters to measure the break avoids the [OIII] and H$\beta$ lines, weaker emission lines contribute to the breaks of approximately half of their objects.

With a break strength of $2.53\pm0.21$, \ruby\ has one of the strongest Balmer breaks measured at $z\gtrsim6$ in any galaxy so far, close to the one of RUBIES-EGS-QG-1 at $z=4.9$ ($2.65\pm0.07$).
In our compilation, only two sources show even stronger breaks: One object from \citet{Wang2024b} which reaches $2.92\pm0.18$ at a slightly lower redshift of $z=6.68$, and the triply imaged LRD from \citet{Furtak2024} with $3.33\pm0.15$ at $z=7.05$. Due to the possible AGN contribution to the spectrum around $4000\,$\AA\ in both of these sources and their likely high dust content, those breaks are however more difficult to interpret \citep[for the latter source, see also][]{Ma2024}.

\subsection{Descendants of \ruby}
\label{sec:descendants}

The early quiescence and high stellar mass of \ruby raises the question of what its descendants may look like. As discussed in Section~\ref{sec:structure}, the high stellar mass surface density is consistent with the range in densities found in compact massive quiescent galaxies at $z\sim2$ \citep[e.g.][]{Whitaker2017}, and also the sources observed with JWST at $z\sim3-5$ \citep{Setton2024,Carnall2023b,deGraaff2024}. Moreover, the 3D stellar mass density profile falls within the $1\sigma$ scatter of the mass profiles of massive quiescent galaxies at $z\sim2-3$ from 3DHST and lies only a factor $\sim2-3$ above the typical profiles of local quiescent galaxies from the GAMA survey, consistent with findings in \citealt{Bezanson2009} where massive quiescent galaxies at $z\sim2-3$ are identified as plausible progenitors of the cores of local ellipticals. This suggests that \ruby is a likely progenitor of the compact massive galaxies seen at $z\sim2-3$, and that the cores of some elliptical galaxies in the local Universe were already in place by $z\sim7$. We note that the inferred metallicity in our low metallicity fit is probably inconsistent with such a scenario which may in turn point towards a higher metallicity for \ruby. As discussed above, accurately constraining $Z_*$ is not possible given the available data and models.

Turning to the SFH, it is tempting to also consider \ruby as a direct progenitor of the extremely old massive quiescent galaxies that have been recently discovered and characterized with JWST at $z\sim2-5$ \citep[e.g.][]{Carnall2024, deGraaff2024,Glazebrook2023,Park2024}: the SFHs of these systems are consistent with having formed their stellar mass at $z\gtrsim8$ and quenched at a similar redshift as \ruby.
However, the stellar masses of these galaxies in the literature are up to an order of magnitude higher. If \ruby were to evolve into such a galaxy, it would either have to rejuvenate quickly and go through a second burst of star formation, to then quench rapidly again. Or alternatively, one or even a few major mergers with similarly massive quiescent systems would be required to reach $\sim10^{11}{\rm M}_\odot$ by $z\sim4-5$. Given the estimated low number density of such objects (see Section \ref{sec:simcomp}), this seems unlikely. A more in-depth analysis of a possible overdensity around \ruby remains to be done in future work. Perhaps most likely, \ruby is not a direct progenitor of the very massive systems at $z\sim4-5$, and may instead maintain its current mass or experience more gradual mass growth to become a massive compact quiescent galaxy as observed at $z\sim2-3$. Interestingly, quiescent galaxies with masses comparable to that of \ruby appear to be the most compact of all quiescent galaxies at $1<z<3$ in \citet{Cutler2024}.

\subsection{Progenitors of \ruby}
\label{sec:progenitors}

Independent of our priors on the SFH and the metallicity in the SED fitting, we know with confidence that \ruby must have formed a substantial fraction of its stars in a burst between $z\sim8-9$ at a star formation rate of $\rm\sim100\,{\rm M}_\odot\,{\rm yr}^{-1}$. Assuming no dust and the conversion from \citet{Kennicutt1998}, this corresponds to a very bright absolute UV-magnitude of $M_{\rm UV}\sim-23.7$. Although a few photometrically-selected candidates exist at $z\sim8$ with $M_{\rm UV}\sim-23$ \citep{Bouwens2022b}, no such object has been spectroscopically confirmed to date, with the brightest spectroscopically-confirmed objects lying around $M_{\rm UV}\sim-22$ at $z\sim8$ \citep[as shown in Figure \ref{fig:sedfit_phot_sfh}; e.g.][]{Roberts-Borsani2024b}. Moreover, these photometric candidates were selected over a very large area of $\sim7\,$deg$^2$ (compared to the $\sim 300\,$arcmin$^2$ survey area of CEERS and PRIMER), and thus have a number density that is nearly two dex lower than our estimate of the number density of \ruby (see Section~\ref{sec:simcomp}). 
Purely star-forming galaxies reaching very bright UV magnitudes ($M_{\rm UV}\sim-23$ to $-24.7$) have recently been found at $z \sim 2.4-3.6$ \citep{Marques-Chaves2020,Marques-Chaves2022}, showing that such objects exist at later times. If and how they are related to \ruby and other high-redshift populations, remains to be established. 

There are a few possible explanations for the missing UV-bright progenitors. First, they may be dust-obscured and thus have so far eluded spectroscopic confirmation by JWST. If they are sufficiently dusty, they may have been challenging to detect prior to JWST and are only recently being followed-up spectroscopically through surveys like RUBIES. One photometric candidate for a dust-obscured progenitor, COS-z8M1 \citep{Akins2023}, is shown in Figure \ref{fig:sedfit_phot_sfh}. An even more extreme candidate at a slightly lower redshift of $z\sim7.6$ is presented in the same paper. Other possible progenitors may be found among the population of LRD-like sources at $z\gtrsim8$, the stellar masses and SFHs of which appear to match well with that of massive quiescent galaxies at $z\sim4-5$ and \ruby \citep{Williams2024,Wang2024b}, although the inferred stellar population properties of these sources are still very uncertain.
However, if these sources do host a significant stellar component, they may evolve into an object like \ruby by $z\sim7$ once the AGN component has shut down and possibly also quenched star formation. It is important to note that the low dust attenuation inferred from the spectrum of \ruby requires that in such a scenario, most of the dust must be destroyed or removed from the galaxy, at least along the line of sight, in a relatively short timescale ($\lesssim100\,$Myr).

Second, the SFH of objects like \ruby could be more extended than the \prospector fits suggest, lowering the required SFR in the most recent burst. The continuity prior applied in our SED-fitting \citep{Leja2019} is however already conservative in this regard. Nevertheless, we find that, if we extend the lower boundary of the metallicity prior in our SED-fitting to $Z=0.01\,{\rm Z_\odot}$, the fit converges to a metallicity of $Z=0.04_{-0.02}^{+0.03}\,{\rm Z_\odot}$ and a more extended SFH that requires substantial star formation ($SFR\sim50\,\Msun\,{\rm yr}^{-1}$) at $z\gtrsim15$. However, these results seem implausible: the very low metallicity lies nearly two dex below the stellar mass--metallicity relation of massive quiescent galaxies at $z\sim1-3$ \citep[e.g.,][]{Choi14,Beverage2024b}, and such an early stellar mass assembly history would likely be in tension with the standard cosmological model \citep{BoylanKolchin2023}. A significantly more extended SFH would also raise the major problem that thus far no sources have been found with such high star formation rates at $z>10$. As can be seen in Figure \ref{fig:sedfit_phot_sfh}, objects as bright as JADES-GS-z14-0, GHZ2 or GN-z11 on the other hand do form plausible progenitors of \ruby for the metallicities explored throughout this paper. 

Third, as discussed in \citet{deGraaff2024} in the context of RUBIES-EGS-QG-1, the number density of sources like \ruby is expected to be low ($\log(n/\rm{Mpc}^{-3})=-5.8^{+0.5}_{-0.8}$, see Section \ref{sec:simcomp}), and uncertain, given that we have only discovered one such source in the available JWST imaging and spectroscopic data. Cosmic variance and the fact that the probed cosmic volume decreases towards higher redshifts may therefore help explaining why we have not yet seen the progenitors of \ruby. Moreover, star formation may happen in short and intense bursts at $z\gtrsim8$ \citep[as suggested by, e.g.,][]{Dekel2023}. Since the observed number density depends not only on the number density of massive galaxies, but also on the duty cycle of star formation, this may explain the lack of observed UV-luminous progenitors at $z>8$. We note that the time of $\sim125\,$Myr that \ruby spends in a bursting phase according to our modeled SFH in e.g., the low metallicity fit (Figure \ref{fig:sedfit_phot_sfh}) is to some extent a result of our choice of SFH time bins. In reality, it may form in one or multiple shorter bursts.
Finally, we note that variations in the initial mass function (IMF) may reduce the inferred stellar mass and thereby the maximum SFR in the SFH \citep[see e.g.,][]{vanDokkum2024}. 

\subsection{The absence of $z>7$ quiescent galaxies in simulations}
\label{sec:simcomp}

\begin{figure}[t!]
     \centering
     \includegraphics[width=0.47\textwidth]{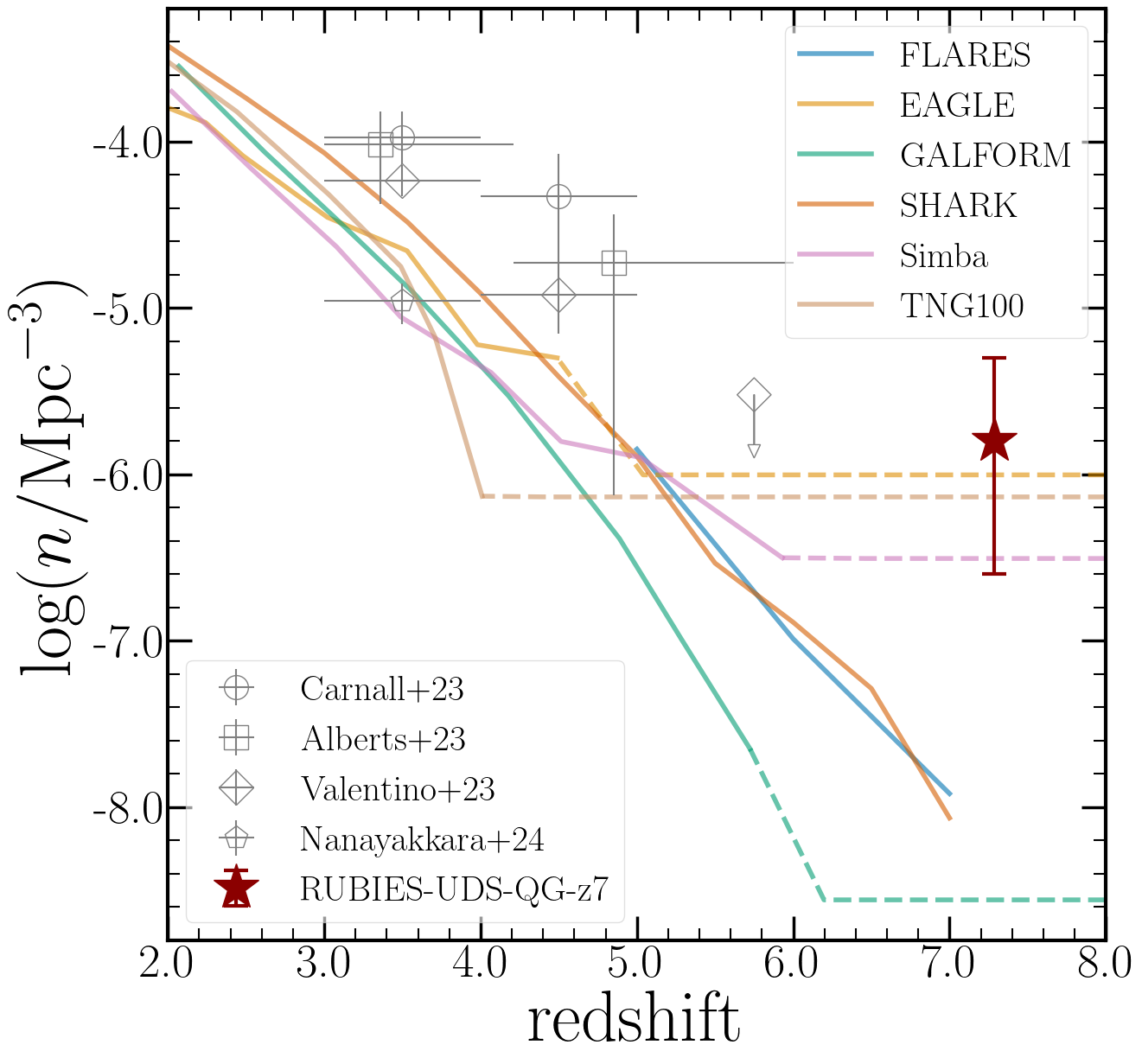}
     \caption{Number density of quiescent galaxies as a function of redshift compared to the inferred number density of objects like \ruby (red star), with $\log(M/\rm{M}_\odot) > 10$ and a specific SFR of $\log(\rm{sSFR}/\rm{Gyr^{-1}})<-1$. Lines are based on various simulations (EAGLE, GALFORM, SHARK, Simba, Illustris TNG100). The dashed lines represent extrapolated upper limits based on the simulation box sizes. Open symbols show lower redshift observational constraints from JWST, where the stellar mass cuts and definitions of quiescence vary between different paper. The highest redshift point from \citet{Valentino2023} is an upper limit, meaning that no quiescent galaxies have been found in the probed volume. The inferred number density of quiescent galaxies at $z\sim7$ based on \ruby in the current survey volume is surprisingly high, and lies $\gtrsim100\times$ above the extrapolation of lower redshift trends.
     }
     \label{fig:sim_comp}
 \end{figure}

Early quiescent galaxies have long represented a key challenge for galaxy formation models and simulations. The challenge is to form a massive galaxy that not only ceases forming stars, but also remains quiescent for a prolonged period of time within the first Gyr of cosmic history. With an inferred stellar mass of $\log(M_*/\rm{M}_\odot) = 10.23\pm0.04$ and a specific SFR of $\log(\rm{sSFR}/\rm{Gyr^{-1}})<-1$ at $z=7.29\pm0.01$, \ruby is by far the most distant galaxy confirmed so far that unambiguously qualifies as a massive quiescent galaxy (see Figure \ref{fig:spectrum_comparison}), pushing the previous record at $z\sim5$ closer to the Big Bang by 500\,Myr. The predicted number density of massive sources ($M_*>10^{10}{\rm M_\odot}$) with $\log(\rm{sSFR}/\rm{Gyr^{-1}})<-1$ drops extremely rapidly with redshift in current galaxy evolution models. For example, the FLARES simulation predicts a number density of $\log(n / \rm{Mpc}^{-3}) < -8$ for such galaxies at $z>7$ \citep{Lovell2023}. 

Assuming that \ruby is unique in the total survey volume of $\sim300$ arcmin$^{2}$ of PRIMER-UDS and CEERS, a conservative estimate of the number density of quiescent galaxies in the redshift bin $z=7-8$ is $\log(n / \rm{Mpc}^{-3}) =-5.8^{+0.5}_{-0.8}$, where uncertainties are computed based on Poisson statistics using the frequentist confidence interval (see \citealt{Maxwell2011}). This is a factor $>150\times$ higher than expected from the FLARES simulation. If we only considered the survey volume of RUBIES alone, i.e., only the area covered by NIRSpec observations \citep[$\sim150$ arcmin$^{2}$;][]{deGraaff2024b}, the inferred number density and the discrepancy with FLARES would be another factor $\sim2\times$ higher.

In Figure \ref{fig:sim_comp}, we compare the number density of $z\sim7$ massive quiescent galaxies (defined here as  sources with $M_*>10^{10}\,\Msun$ and $\log(\rm{sSFR}/\rm{Gyr^{-1}})<-1$) inferred from \ruby to the number densities of such systems measured from various simulations (FLARES, \citealt{Lovell2023}; EAGLE, \citealt{Schaye2015,Crain2015}; GALFORM, \citealt{Lacey16}; SHARK, \citealt{Lagos2018,Lagos2024}, Simba, \citealt{Dave19}; Illustris-TNG100 \citealt{Pillepich2018b}), as well as recent observational constraints based on JWST data \citep{Nanayakkara2024,Carnall2023,Valentino2023,Alberts23}, although we note that these observational studies use different selection criteria for the stellar mass and different definitions of quiescence. Apart from FLARES and SHARK, simulations largely only provide upper limits on the number density of massive quiescent galaxies at $z\gtrsim6$, because there are no such sources within the respective simulated volumes. The steeply declining number densities at $z\sim2-6$ seen in all the simulations shown in Figure \ref{fig:sim_comp} however suggest number densities at $z\sim7$ comparable to or even lower than those in FLARES.

This indicates that star formation efficiencies, cold gas in- and outflows as well as feedback mechanisms may have to be revisited in simulations at high redshifts, in order to grow and subsequently quench more massive galaxies early on. From the observational side, it will be important to improve the number density estimates of sources like \ruby in wider area data covering larger volumes to provide more stringent constraints. \citet{Lagos2024b} also show that even at lower redshifts, $z\approx 3-4$, where some of the simulations do predict enough massive quiescent galaxies, their predicted SFHs may not be bursty enough compared with observations. This shows that measuring number densities, together with inferring SFHs and other intrinsic galaxies properties, provides stringent constraints on cosmological galaxy formation simulations.

\section{Summary and Conclusions}
\label{sec:conclusions}

We have presented the NIRSpec/PRISM spectrum of a massive quiescent galaxy, \ruby, at $z=7.29\pm0.01$ which was observed as part of the JWST Cycle 2 program RUBIES.
\ruby represents the highest redshift massive quiescent galaxy known to date by $\Delta z>2$. Through simultaneous modeling of the spectrum and the NIRCam and MIRI photometry, we find that \ruby formed most of its mass of log$(M_*/M_\odot)=10.23^{+0.04}_{-0.04}$ in a burst of star formation (${\rm SFR_{\rm peak}}\sim100\,{\rm M_\odot}{\rm yr}^{-1}$) at $z\sim8-9$ and then stopped forming stars quickly, resulting in a low $\log(\rm{sSFR/Gyr}^{-1})<-1$ in the last $50\,$Myr.
While the stellar mass and the rough shape of the SFH are well constrained, we find similarly good fits for metallicities $<0.1\,{Z_\odot}$ as well as $>1\,{Z_\odot}$, likely due to non-solar abundance patterns in early and rapidly forming galaxies like \ruby which are not yet accounted for in available stellar population models.

The compact morphology of \ruby ($\re=209^{+33}_{-24}\,{\rm pc}$) implies high stellar mass densities comparable to those measured in massive quiescent galaxies at $z\sim4-5$, the densest quiescent systems at $z\sim2-3$, as well as the inner $\sim300\,$pc of local ellipticals, indicating that the cores of some of them may be in place already at $z\sim7$. 

Progenitors of \ruby are expected to be highly star-forming systems at $z\sim8-9$. Only few photometric candidates with UV-magnitudes directly implying sufficiently high SFRs have been found at those redshifts, and only over much larger areas. However, the progenitors of \ruby may be dust-obscured making their detection and characterization more challenging. Photometric candidates for such dust-obscured, highly star-forming systems at $z\sim8$ are slowly being found thanks to deep NIRCam+MIRI imaging \citep[e.g.,][]{Akins2023}. Spectroscopic confirmation of such candidates will be critical to shed more light on the possible formation pathway of \ruby. 

The detection of \ruby in a survey area of just $\sim300$ arcmin$^2$ implies a number density of $\log(n/\rm{Mpc}^{-3}) =-5.8^{+0.5}_{-0.8}$, which is comparable to observations of quiescent galaxies at $z\sim4-5$, but $>100\times$ higher than predictions by simulations at $z\sim7$. Creating such a distant quiescent galaxy therefore represents a challenge for our current galaxy formation theories and may require a revision of our modeling assumptions.

In the future, it will be critical to search for sources similar to  \ruby at these redshifts, but over wider area JWST fields in order to refine the number density estimate of quiescent galaxies at $z>6$. Additionally, deep medium/high resolution NIRSpec spectroscopy has the potential to reveal various absorption features for \ruby itself, which are only tentatively or not at all detected in the current PRISM spectrum. These would provide better constraints on the SFH and a more direct measurement of elemental abundances. Finally, sub-mm  observations from, e.g., ALMA could constrain the gas and dust properties to gain further insight into the possible quenching mechanisms. \ruby thus represents a unique opportunity to study and understand the emergence of the first quiescent galaxies in the early Universe.

\begin{acknowledgements}
\section*{Acknowledgements}

We thank the PRIMER team for making their imaging data publicly available immediately. 
This work is based on observations made with the NASA/ESA/CSA James Webb Space Telescope. The data were obtained from the Mikulski Archive for Space Telescopes at the Space Telescope Science Institute, which is operated by the Association of Universities for Research in Astronomy, Inc., under NASA contract NAS 5-03127 for JWST. These observations are associated with program \#4233.
Support for program \#4233 was provided by NASA through a grant from the Space Telescope Science Institute, which is operated by the Association of Universities for Research in Astronomy, Inc., under NASA contract NAS 5-03127.
This research was supported by the International Space Science Institute (ISSI) in Bern, through ISSI International Team project \#562.
The Cosmic Dawn Center is funded by the Danish National Research Foundation (DNRF140). This work has received funding from the Swiss State Secretariat for Education, Research and Innovation (SERI) under contract number MB22.00072, as well as from the Swiss National Science Foundation (SNSF) through project grant 200020\_207349. Support for this work was provided by The Brinson Foundation through a Brinson Prize Fellowship grant.
Support for this work for RPN was provided by NASA through the NASA Hubble Fellowship grant HST-HF2-51515.001-A awarded by the Space Telescope Science Institute, which is operated by the Association of Universities for Research in Astronomy, Incorporated, under NASA contract NAS5-26555. 

TBM was supported by a CIERA fellowship.





\facilities{JWST(NIRSpec, NIRCam)}

\software{
All software packages used in this work  are publicly available on Github: \texttt{grizli}, \texttt{msafit}, \texttt{msaexp}, \prospector, \texttt{sedpy}. We acknowledge: 
    astropy \citep{2013A&A...558A..33A,2018AJ....156..123A,2022ApJ...935..167A}, 
    matplotlib \citep{10.1109/MCSE.2007.55},  
    numpy \citep{10.1038/s41586-020-2649-2}, 
    scipy \citep{10.1038/s41592-019-0686-2}, 
    lmfit \citep{matt_newville_2024_12785036},
    eMPT \citep{2023A&A...672A..40B}, 
    the \texttt{jwst} pipeline (\citealt{10.5281/zenodo.10870758}), 
    \texttt{msaexp} (\citealt{10.5281/zenodo.7299500}), 
    \texttt{grizli} (\citealt{10.5281/zenodo.1146904}),  
}

\end{acknowledgements}

\bibliography{paper}{}

\bibliographystyle{aasjournal}

\appendix

\section{G395M Spectrum}
\label{appendix:grating_spectrum}

\begin{figure*}[htp!]
     \begin{center}
     \includegraphics[width=0.9\textwidth]{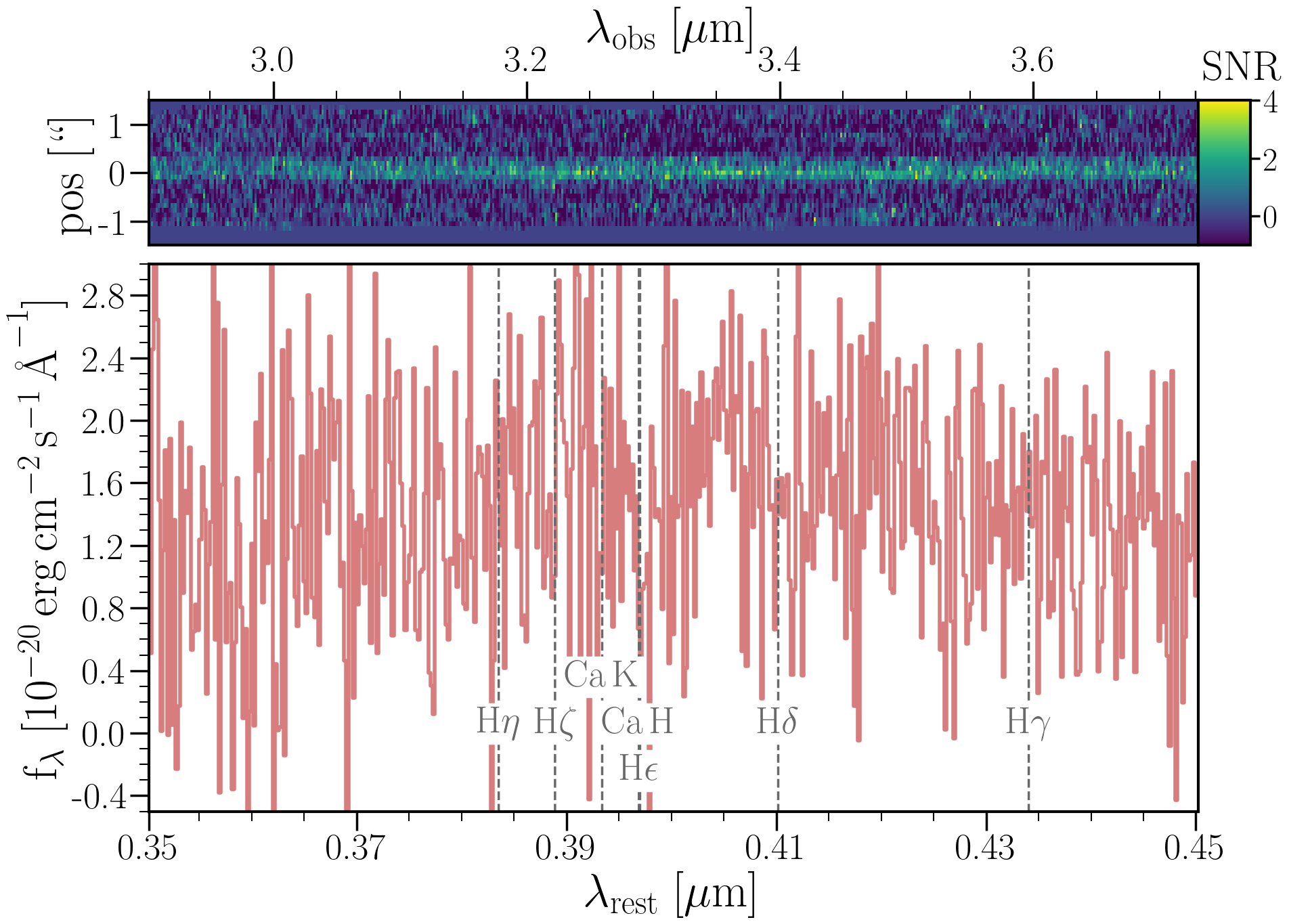}
     \end{center}
     \caption{NIRSpec/G395M spectrum of \ruby, zoomed in on the region around $0.4\,\mu m$. The position of possible absorption features are highlighted, but the SNR of the spectrum is not sufficient to unambiguously reveal any significant features.}
     \label{fig:spectrum_g395}
\end{figure*}

In addition to the PRISM spectrum shown above, RUBIES also obtained a higher resolution G395M grating spectrum. This is shown in Figure \ref{fig:spectrum_g395}, zoomed-in on the $\lambda_{\rm rest}\in(0.35,0.45)\,\mu m$ range. The SNR is not sufficient to reveal any absorption features or provide any other further insights, which is why we do not use the grating spectrum in the analysis outlined in this work.

\section{\prospector Posterior Distributions}
\label{appendix:posterior}

\begin{figure*}[htp!]
     \begin{center}
     \includegraphics[width=0.9\textwidth]{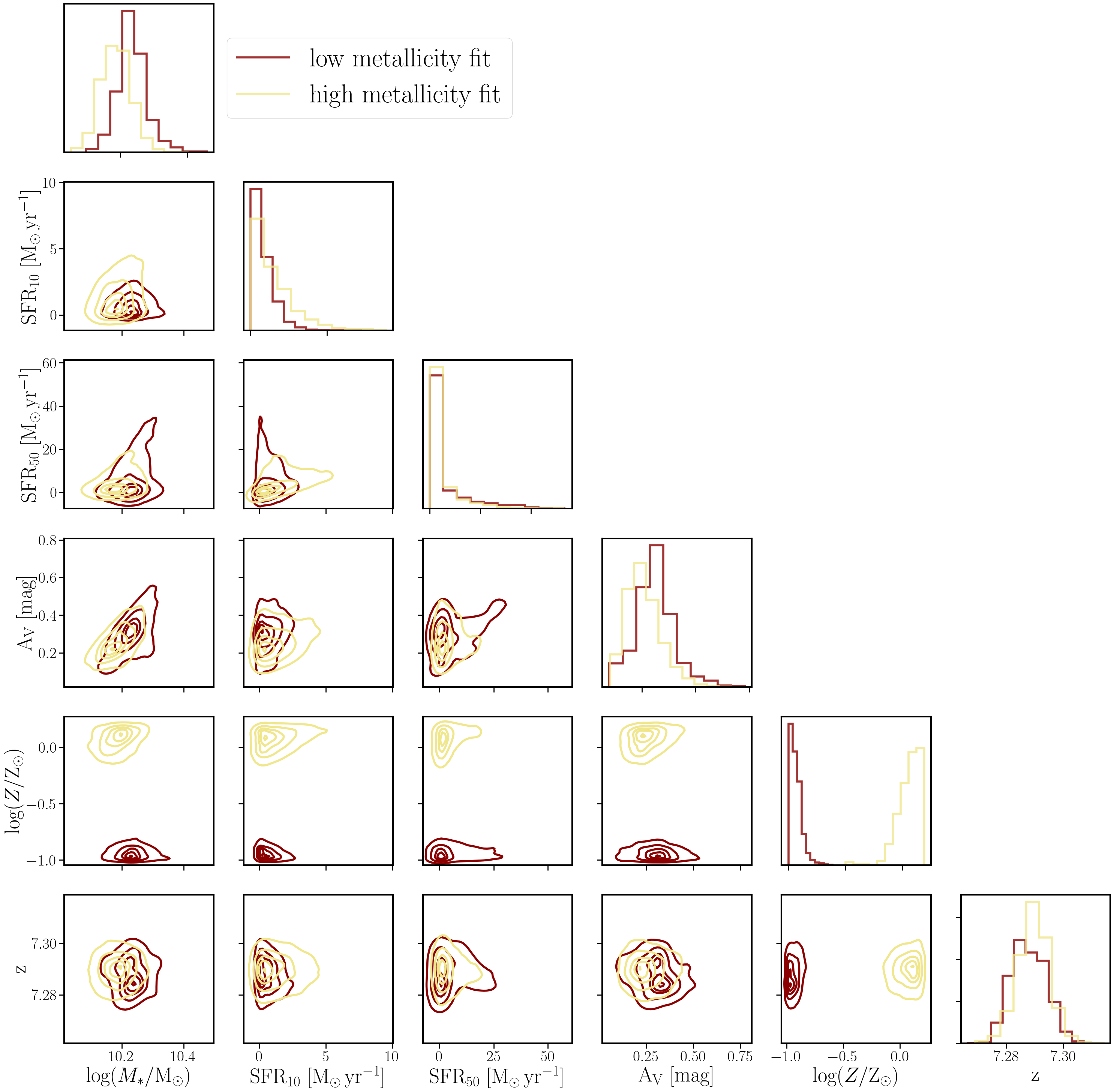}
     \end{center}
     \caption{Posterior distributions of the key parameters from the \prospector runs. Results from the low metallicity fit are shown in red, and those from the high metallicity run in khaki. The displayed contours correspond to 0.5, 1, 1.5, 2, and 3$\sigma$ confidence regions.}
     \label{fig:posterior}
\end{figure*}

As an overview of the SED-fitting results from our \prospector runs, we present the posterior distributions of the key parameters in a corner plot in Figure \ref{fig:posterior}.

\end{document}